\newif\ifanonymous
\anonymoustrue

\documentclass[11pt]{scrartcl}
\usepackage{authblk}
\usepackage[utf8]{inputenc}
\usepackage[]{algorithm2e}
\usepackage{amssymb,amsmath}
\usepackage{color}
\usepackage{geometry}
\usepackage{graphicx}
\usepackage{tabularx}
\usepackage{subfig}

\usepackage[scaled=.95]{helvet}
\usepackage{courier}

\graphicspath{{./}{figures/}}

\usepackage[round]{natbib}

\usepackage{tikz}
\usetikzlibrary[positioning]
\usetikzlibrary{patterns}

\newcommand{\du}{\,\mathrm{d}}
\newcommand{\R}{\mathbb{R}}

\newcommand{\im}{\mathrm{i}}

\newcommand{\xb}{\mathbf{x}}
\newcommand{\vb}{\mathbf{v}}
\newcommand{\wb}{\mathbf{w}}

\newcommand{\Eb}{\mathbf{E}}
\newcommand{\Bb}{\mathbf{B}}

\begin{document}

\title{A massively parallel semi-Lagrangian solver for the six-dimensional
Vlasov--Poisson equation\footnote{K.K.~and K.R.~contributed equally to this work.}}

\author[1]{Katharina Kormann}
\author[2]{Klaus Reuter}
\author[2]{Markus Rampp}
\affil[1]{Max-Planck-Institut für Plasmaphysik \& \newline Technische Universit\"at M\"unchen}
\affil[2]{Max Planck Computing and Data Facility}
\maketitle




\begin{abstract}
This paper presents an optimized and scalable semi-Lagrangian solver for the
Vlasov--Poisson system in six-dimensional phase space.  Grid-based solvers of
the Vlasov equation are known to give accurate results.  At the same time, these
solvers are challenged by the curse of dimensionality resulting in very high
memory requirements, and moreover, requiring highly efficient parallelization
schemes.
In this paper, we consider the 6d Vlasov--Poisson problem discretized by a
split-step semi-Lagrangian scheme, using successive 1d interpolations on 1d
stripes of the 6d domain. Two parallelization paradigms are compared, a
remapping scheme and a classical domain decomposition approach applied to the
full 6d problem. From numerical experiments, the latter approach is found to be
superior in the massively parallel case in various respects.
We address the challenge of artificial time step restrictions due to the
decomposition of the domain by introducing a blocked one-sided communication
scheme for the purely electrostatic case and a rotating mesh for the case with a
constant magnetic field. In addition, we propose a pipelining scheme that
enables to hide the costs for the halo communication between neighbor processes
efficiently behind useful computation.
Parallel scalability on up to 65k processes is demonstrated for benchmark
problems on a supercomputer.
\end{abstract}


\section{Introduction}

Numerical simulations are of key importance for the understanding of the
behavior of plasmas in a nuclear fusion device.  The fundamental model in plasma
physics is a kinetic description by a phase space distribution function solving
the Vlasov--Maxwell equation.

State-of-the-art kinetic simulations for magnetic confinement fusion are built
upon the so-called gyrokinetic model, a reduced model in a five-dimensional
phase space. Since the increase in parallel computing power renders the solution
of the fully kinetic Vlasov equation in six-dimensional phase space possible,
interest in solving the 6d Vlasov equation has arisen recently
(cf.~e.g.~\citet{Munoz15, Groselj17, Kuley15, Miecnikowski18}). For the solution of the Vlasov equation, both
grid-based and particle-based methods are commonly used.  We distinguish
two classes of grid-based methods, Eulerian solvers based on finite volume or
discontinuous Galerkin on the one hand, and, on the other hand, semi-Lagrangian
methods that update the solution by evolution along the characteristics using
interpolation. The latter class of methods has the advantage that usually no
time step restriction by a Courant--Friedrich--Levy (CFL) condition needs to be
imposed.
Recently, progress has been made to develop efficient implementations and
algorithms for the solution of the Vlasov--Poisson problem with a
particle-in-cell method on GPUs \citep{Hariri16}, semi-Lagrangian methods based
on compressed grids \citep{Kormann15, Kormann16, Guo16, Einkemmer18} and finite
differences on adaptive grids \citep{Deriaz15}.  Solvers for the
six-dimensional Vlasov equation have also been developed in the space plasma
community. In particular, Yoshikawa, Yoshida, Umemura and coworkers have
presented Vlasov--Poisson and Vlasov--Maxwell solvers in six-dimensional phase
space based on semi-Lagrangian methods \citep{Tanaka17, Yoshikawa13}. Another
example is the hybrid Vlasov--Maxwell (HVM) code based on finite differences
\citep{Mangeney02, Cerri18}.

Typically particle-in-cell methods are used for high-dimensional simulations due
to their favorable scaling with the dimensionality. Moreover, the particle pusher
part is embarrassingly parallel. On the other hand, particle-in-cell methods
suffer from numerical noise.
Due to the high memory footprint of six
dimensional grids, an efficient parallelization that scales to high-performance
computer systems is essential for the success of grid-based algorithms for the
solution of a fully kinetic description of a plasma. In this paper, we focus on
efficient parallelization schemes for a semi-Lagrangian discretization of the
Vlasov--Poisson equation in six-dimensional phase space (neglecting magnetic
effects). For a Vlasov--Poisson equation on a four-dimensional phase space, two
parallelization schemes have been discussed in the literature: a domain
partitioning scheme with patches of four-dimensional data blocks
\citep{Crouseilles09} as well as a remapping scheme \citep{Coulaud99}. The idea of
the remapping scheme is to work with two different domain partitions which both
keep a partition of the dimensions sequential on each processor. While the
latter strategy is very well adapted to a semi-Lagrangian method combined with
dimensional splitting, its parallel scalability is hampered by an all-to-all
communication pattern. For domain decomposition schemes, typically time step
restrictions need to be introduced since the interpolation stencil needs to be
localized at the boundaries of the computational domains (cf. \citet[Sec.
2.4]{Yoshikawa13} and \citet{Crouseilles09}). This restriction is particularly
severe for magnetized plasmas in a strong guide field where particles perform a
fast gyromotion around the magnetic field lines causing strong non-locality in
the velocity domain. In this paper, we therefore propose the use of a rotating
velocity grid.
Compared to three-dimensional problems where domain decomposition schemes are
very successful a domain decomposition in six dimensions requires a much larger
degree of communication due to the fact that the surface-to-volume ratio of a $d$
dimensional data block increases with the dimension $d$. For a gyrokinetic code
on the contrary one has the additional advantage that the fifth dimension of the
model is merely a parameter in the gyrokinetic Vlasov equation so that
communication in the Vlasov step is reduced to four dimensions. This can be
particularly advantageous when parallelization over several islands of a
supercomputer are considered: If the parameter dimension is the only one split
between islands, the comparably slow communication over island barriers can be
avoided. Good scaling properties for a semi-Lagrangian solver of the gyrokinetic
model has for instance been demonstrated for the GYSELA code by
\citet{Grandgirard16} and \citet{Latu16}.  Another challenge compared to the gyrokinetic
model is the reduced complexity of the kinetic model causing a reduced
computational complexity that renders the code more memory-bound. In order to
reduce the burden of the high demand on MPI-data-exchange, we propose a blocking
scheme to overlap communication and computations in the advection steps.




The outline of the paper is as follows: In the next section, we introduce the
physical model, the semi-Lagrangian method including the rotating mesh for the background magnetic field, and the parallelization schemes.
Moreover, we discuss the impact of the parallelization on the interpolation step
in the semi-Lagrangian scheme in Sec.~\ref{sec:interpolation}. In Sec.~\ref{sec:implementation}, we describe our
implementation of the domain partitioning scheme, followed by a discussion of our performance optimizations in Sec.~\ref{sec:optimizations}. Sec.~\ref{sec:numerics} compares Lagrange interpolation of various orders and demonstrates the effect of the rotating grid followed by a numerical
demonstration of the scalability of our new implementation in
Sec.~\ref{sec:performance}. Finally, Sec.~\ref{sec:conclusions} concludes the
paper.

\section{Algorithmic background}

\subsection{Vlasov--Poisson equation}

The Vlasov--Poisson equation describes the motion of a plasma in its
self-consistent electric field for low-frequency phenomena.  The Vlasov equation
for electrons in dimensionless form is given as
\begin{equation*}
\begin{aligned}
\partial_t f(\xb,\vb,t)
&+ \vb \cdot \nabla_{\xb} f(\xb,\vb,t) \\
&- \left(\mathbf{E} + \vb \times \Bb_0 \right) ( \xb,t) \cdot \nabla_{\vb} f(\xb,\vb,t) = 0.
\end{aligned}
\end{equation*}
The self-consistent field for electrons in a neutralizing ion background can be
computed by the following Poisson equation,
\begin{equation}\label{eq:poisson}\begin{aligned}
&-\Delta \phi(\xb,t) = 1-\rho(\xb,t), \quad \mathbf{E}(\xb,t)=-\nabla \phi(\xb,t), \\
&\quad \rho(\xb,t)=\int f(\xb,\vb,t) \du \vb.
\end{aligned}\end{equation}
Here, $f$ denotes the probability density of a particle in phase space defined
by position $\xb \in D \subset \R^3$ and velocity $\vb \in \R^3$, $\mathbf{E}$ denotes
the electric field, $\phi$ the electric potential, and $\rho$ the charge
density. The magnetic field $\Bb_0$ is supposed to be either zero or a constant background field aligned with the $x_3$ axis.
Generally, the spatial domain is defined by the geometry of a tokamak
or similar fusion device.
In this paper, we restrict ourselves to common benchmark problems on a
periodic box. The distribution has a Maxwellian shape in velocity such that it
follows an exponential decay for large values of the velocity. We therefore
truncate the computational domain in velocity space to a box and close the
system with (artificial) periodic boundary conditions.

The characteristic curves of the Vlasov equation can be defined by the following
system of ordinary differential equations (ODE),
\begin{equation}\label{eq:characteristics}
\frac{\du \mathbf{X}}{\du t} = \mathbf{V}, \quad \frac{\du \mathbf{V}}{\du t} =
-\left(\mathbf{E}(\mathbf{X},t) + \mathbf{V} \times \Bb_0 \right).
\end{equation}
Let us denote by $\mathbf{X}(t;\xb,\vb,s), \mathbf{V}(t;\xb,\vb,s)$ the solution
of the characteristic equations \eqref{eq:characteristics} at time $t$ with
initial conditions $\mathbf{X}(s)=\xb$ and $\mathbf{V}(s) = \vb$. Given an
initial distribution $f_0$ at time $t_0$, the solution at time $s>0$ is given by
\begin{equation}\label{eq:f_from_characteristics_and_f0}
f(\xb,\vb,s) = f_0(\mathbf{X}(t_0;\xb,\vb,s),\mathbf{V}(t_0;\xb,\vb,s)),
\end{equation}
since the distribution function is constant along the characteristic curves.


\subsection{The semi-Lagrangian method for Vlasov--Poisson}\label{sec:sl}

To numerically compute the solution of the Vlasov equation, we use the so-called
semi-Lagrangian method. We introduce a six-dimensional grid to discretize the
phase space. In each time step, the equations for the characteristics are solved
for each grid point backwards in time from time $t_{m+1}$ to time $t_m$ with
$\Delta t = t_{m+1} - t_m$ small. Then
equation \eqref{eq:f_from_characteristics_and_f0} is used with $s=t_{m+1}$ and
$t_0=t_m$ to find the solution at time $t_{m+1}$ for each grid point. Since
$f^{(m)}$ is only known on the grid points, some interpolation method is needed
to approximate the value of
$f^{(m)}(\mathbf{X}(t_m;\xb,\vb,t_{m+1}),\mathbf{V}(t_m;\xb,\vb,t_{m+1}))$.
In this general form, the solution with a semi-Lagrangian method requires the
solution of a system of ODE as well as interpolation.

To efficiently solve the characteristic equations, \citet{Cheng76}
proposed a splitting method for the Vlasov--Poisson equation ($\Bb_0=0$) reduced to a 2d (1x--1v)
phase space that splits the $\xb$ and $\vb$ advections. This yields the
following algorithm: Given $f^{(m)}$ and $\Eb^{(m)}$ at time $t_m$, we compute
$f^{(m+1)}$ at time $t_m + \Delta t$ for all grid points $(\xb_i, \vb_j)$ as
follows:
\begin{enumerate}
  \item Solve $\partial_t f - \Eb^{(m)} \cdot \nabla_{\vb} f = 0$ on half time
        step:\newline
        $f^{(m,*)}(\xb_i,\vb_j) = f^{(m)}(\xb_i,\vb_j+\Eb_i^{(m)}\frac{\Delta t}{2})$
  \item Solve $\partial_t f + \vb \cdot \nabla_{\xb} f = 0$ on full time
        step:\newline
        $f^{(m,**)}(\xb_i,\vb_j) = f^{(m,*)}(\xb_i-\vb_j\Delta t,\vb_j)$
  \item Compute $\rho(\xb_i,\vb_i)$ and solve the Poisson equation for $\Eb^{(m+1)}$
  \item Solve $\partial_t f - \Eb^{(m+1)} \cdot \nabla_{\xb} f = 0$ on half time
        step:\newline
        $f^{(m+1)}(\xb_i,\vb_j) = f^{(m,**)}(\xb_i,\vb_j+\Eb_i^{(m+1)}\frac{\Delta t}{2})$
\end{enumerate}
Note that the electric field is constant for the $\vb$ advection step.
Therefore, the advection coefficients are independent of $\vb$ for the $\vb$
advection and independent of $\xb$ for the $\xb$ advection and the
characteristics are therefore given analytically.

In order to avoid three-dimensional interpolation, we use a cascade
interpolation scheme replacing the three-dimensional interpolation by three
successive one-dimensional interpolations.
Moreover, we can use a first-same-as-last implementation that clusters step 4 of
time step $m$ with step 1 of time step $m+1$.

As a consequence the main building block of the split-step semi-Lagrangian
discretization of the Vlasov--Poisson problem is one-dimensional interpolation
on one-dimensional stripes of the six-dimensional domain.
Moreover, the interpolation step on the individual stripes has a special form:
The function needs to be interpolated at a shifted value of each grid point and
the value of this shift is constant for the whole stripe.  For a stripe of
length $n$ with grid points $x_i$, $i=1,\ldots, n$, we compute
\begin{equation}\label{eq:1d_interpolation}
  g^{(m+1)}(x_i) = g^{m}(x_i + \alpha), \quad i=1, \ldots, n.
\end{equation}
Since $\alpha$ is independent of $x_i$, the interpolation formula is the same in
each grid cell which can be exploited for vectorized implementation.

Advections can also be reduced to one-dimensional interpolation for the
Vlasov--Maxwell equation using the backward substitution method introduced by
\citet{Schmitz06}. In this paper, we focus on the Vlasov--Poisson problem.
However, we include strong background magnetic fields and discuss in the next
section how they can be integrated into the split-step semi-Lagrangian scheme.

\subsection{Split-step semi-Lagrangian method on a rotating mesh}

In a magnetic confinement fusion device, the background magnetic field is strong
compared to the self-consistent fields and causes a rapid motion around the
field lines, the so-called gyromotion. Often the time scale of the gyromotion is
the fastest so that we do not want to accurately resolve this time scale. On the other hand, the rotation around the magnetic axis (here the $x_3$ axis) causes non-locality in the perpendicular velocity plane (the $(v_1,v_2)$ plane) which is difficult to handle when working with distributed grids.

We therefore use a rotating grid that follows the circular motion of the characteristics given by
\begin{equation}
\frac{\du \mathbf{V}}{\du t} = \mathbf{V} \times \Bb_0,
\end{equation}
Moving grids for Vlasov simulations have previously been discussed by
\citet{Sonnendrucker04}.

To this end, we define a logical grid equivalent to the physical grid at initial
time. Let us define the rotation matrix
\begin{equation}
D(t) =
\begin{pmatrix}
  \cos(B (t-t_0)) & \sin( B (t-t_0)) & 0 \\
  -\sin(B (t-t_0)) & \cos( B (t-t_0)) & 0 \\
  0 & 0 & 1 \\
\end{pmatrix},
\end{equation}
where $\Bb_0 = B \hat{\mathbf{x}}_3$. The logical grid then follows the fast gyromotion with rotation frequency $\omega_c = \frac{2\pi}{B}$.

For a velocity $\wb$ on the logical grid, the physical value of the velocity at time $t_m$ is given by $\vb (\wb)  = D(t_m) \wb$. Furthermore, let us denote by $f^{(m)}$ the distribution function on the physical grid at time $t_m$ and by $g^{(m)}$ the distribution function on the logical grid at time $t_m$, i.e.
\begin{equation}
  f^{(m)}(\xb,D(t_m)\wb ) = g^{(m)}(\xb, \wb).
\end{equation}
In the advection steps, we always solve the characteristic equations in physical coordinates and then transform the resulting velocity coordinates to the logical grid.

We again split the $\xb$ and $\vb$ advections.
Then, the solution at time $t$ of the separate characteristic equations starting
at $(\xb,\vb)$ at time $t_0$ is given as
\begin{equation}
X_i(t;t_0,x,v) = x_i + (t-t_0)v_i, \; i=1,2,3,
\end{equation}
\begin{equation}\label{eq:char_v}
{\small
\begin{aligned}
&\textbf{V}(t;t_0,v) = \begin{pmatrix}
\cos((t-t_0) B) & \sin((t-t_0) B) & 0 \\ -\sin((t-t_0) B) & \cos((t-t_0) B) & 0 \\ 0 & 0 & 1
\end{pmatrix} v \\
&+\frac{1}{B} \begin{pmatrix}
\sin((t-t_0) B) & 1-\cos((t-t_0) B) & 0 \\ \cos((t-t_0) B)-1 & \sin((t-t_0) B) & 0 \\ 0 & 0 & B(t-t_0)
\end{pmatrix} \\
&\Eb(\xb, t).
\end{aligned}
}
\end{equation}

This yields the following advection steps on the rotating grid:
\paragraph{$\vb$ advection:} Defining
\begin{equation}
\begin{aligned}
&A(t) = \frac{1}{B} \times \\
&\begin{pmatrix}
\sin((t-t_0) B) & 1-\cos((t-t_0) B) & 0 \\ \cos((t-t_0) B)-1 & \sin((t-t_0) B) & 0 \\ 0 & 0 & B(t-t_0)
\end{pmatrix}
\end{aligned}
\end{equation}
for given $\vb$ at time $t_{m+1}$ the origin of the characteristic \eqref{eq:char_v} at time $t_m$ is given by $\mathbf{V}(t_m;t_{m+1},v) = D(t_m-t_{m+1}) v + A(t_m-t_{m+1}) \Eb =  D(t_{m+1}-t_m)^{-1} v + A(t_m-t_{m+1}) \Eb $. For the $\vb$ advection, we work with different physical grids at time $t_m$ and $t_{m+1}$, namely
\begin{equation}
\begin{aligned}
  &f^{(m)}(\xb,D(t_m)\wb ) = g^{(m)}(\xb, \wb), \quad \\
  &f^{(m+1)}(\xb,D(t_{m+1})\wb ) = g^{(m+1)}(\xb, \wb).
\end{aligned}
\end{equation}

To find the representation of $g^{(m+1)}$ at a point $\wb$ of the logical grid, we first transform to the representation on the physical grid, use the characteristic equation to express it in terms of the solution at time $t_m$ and finally transform back to the logical grid at time $t_m$:
\begin{equation}
\begin{aligned}
&g^{(m+1)}(\xb, \wb) = f^{(m+1)}(\xb , D(t_{m+1})\wb) \\
&= f^{(m)}(x,D(t_{m+1}-t_m )^{-1} D (t_{m+1}) \wb + A \Eb) \\
&= g^{(m)} \left(\xb, D(t_m)^{-1} \left( D(t_m) \wb + A \Eb \right)\right) \\
&= g^{(m)} (\xb, \wb + D(t_m)^{-1} A \Eb).
\end{aligned}
\end{equation}
Note that the displacement $D(t_m)^{-1} A \Eb$ on the logical grid is not dependent on $\wb$. In our implementation, we compute $D(t_m)^{-1} A \Eb$ and then use successive one-dimensional interpolations along the three velocity coordinates axes of the logical grid.
\paragraph{$\xb$ advection:}
In this step, the transformation between the physical and the logical grid does not change. We therefore have
\begin{equation}
\begin{aligned}
  g^{(m+1)}(\xb,\wb) &= f^{(m+1)} (\xb, D(t_m)\wb) \\
    &= f^{(m)}(\xb - \Delta t D(t_m) \wb, D(t_m)\wb) \\
    &= g^{(m)}(\xb - \Delta t D(t_m) \wb,\wb).
\end{aligned}
\end{equation}
Compared to the case with fixed grid, the displacements in $x_1$ and $x_2$ are now dependent on both $v_1$ and $v_2$ and on time. On the other hand, we still have displacements independent of $\xb$ and we can use successive 1d interpolations along the three coordinate axes.

\subsection{Domain partitioning}

Due to the curse of dimensionality, the memory requirements of a grid in the six
dimensional phase space are rather high already for coarse resolutions.
Therefore, a distributed numerical solution of the problem is inevitable. Two
choices of domain partitioning are considered in this paper:

\begin{itemize}
\item Domain decomposition \citep{Crouseilles09}: The domain is decomposed into
patches of 6d data blocks, each representing a separate part of the domain. The
patches are mapped to a 6d logical grid of processors.  For interpolations
next to the domain boundary, halo cells with a width determined by the
interpolation stencil have to be introduced and filled in advance with data
from neighboring processors.  This classical approach is widely used in
parallelizations of lower-dimensional physics and engineering problems, e.g.
2d or 3d computational fluid dynamics.
\item Remap \citep{Coulaud99}: Two decompositions of the domain are introduced,
the first one distributing the domain only over the velocity dimensions
(keeping the spatial dimensions local to each processor) and the second one
distributing the domain only over the spatial dimensions (keeping the velocity
dimensions local to each processor). For $\xb$ advection steps, we use the
first decomposition and for the $\vb$ advection steps the second. In between
the steps, the data is redistributed between the two decompositions using an
all-to-all communication pattern.
\end{itemize}

The first strategy has the clear advantage over the remap method that the
complexity of the communication pattern is reduced dramatically.  On the other
hand, the remap scheme is very well adapted to the split-step semi-Lagrangian
method since the one-dimensional interpolations can be performed locally once
the remapping has been applied. For the domain decomposition method the one
dimensional stripes are usually distributed over separate domains. This makes
the implementation more complicated and introduces an artificial time-step
restriction (similar to a CFL condition) since the interpolation needs
information of the function around the shifted data point $x_i+\alpha$ in
\eqref{eq:1d_interpolation}.

\subsection{Solution of Poisson's equation}

The focus of this article is on the distributed solution of the 6d Vlasov
equation. However, in addition we need to solve the 3d Poisson problem
\eqref{eq:poisson}. Since the problem is only three-dimensional, the compute
time spent on its solution is almost negligible. For this reason, we use a
pseudo-spectral solver based on fast Fourier transforms (FFTs) for the solution
of the Poisson equation and remap the solution between domain decompositions
that are local along the direction where FFTs are performed. In case a full 6d
domain decomposition is used (i.e. when the widths of all dimensions of
the logical grid of processors are greater than 1), there are several subgroups
of processors that span the whole $\xb$ or $\vb$ domain, respectively.  As a
first step, we compute the charge density by a reduction along the velocity
dimension. This involves an all-to-all communication among groups of MPI
processes of equal spatial domain. Then, the Poisson equation is solved on each
subgroup of processors that span the whole $\xb$ domain. By solving the same
Poisson problem in each subgroup, we avoid another communication step for
redistribution of the computed electric field.

Finally, we also include a diagnostic step in our stimulations that computes
scalar quantities like mass, momentum, and energy, thus containing
reduction steps over the full 6d array.

\section{Interpolation on distributed domains}\label{sec:interpolation}


To compute the interpolated values, we can either use nodal interpolation
formulas like Lagrange interpolation or global interpolants like spline
interpolation. For simulations of the Vlasov--Poisson problem, cubic spline
interpolation is most popular since it is well balanced between accuracy and
efficiency. In combination with a domain decomposition, we however have to deal
with the fact that the stripes are distributed between several processors,
rendering a global interpolant impractical due to the required data exchange.
Local splines as e.g.~considered for the 4D Vlasov--Poisson equation by
\citet{Crouseilles09} are an interesting alternative. However, in this paper, we
focus on local Lagrange interpolation.

Let us recall the special structure of the interpolation task arising from our 1d
advections: The new value at grid point $x_i$ is given by the interpolated value
at $x_i+\alpha$ for some displacement $\alpha$ that is constant over the whole
stripe. Let us decompose the displacement $\alpha$ into a multiple $\gamma \in
\mathbb{Z}$ of $\Delta x$ and a remainder $\beta \in [0,1]$, i.e.
\begin{equation}
\alpha = \left( \gamma + \beta\right) \Delta x.
\end{equation}
Depending on the sign of $\alpha$, the origin of the characteristics for points
close to the boundary of the local domain are displaced into a part of the
domain that is stored on a neighboring process. Since the interpolation formula
needs to be centered around $x_i+\alpha$ this yields an additional need for halo
data points on one side of the domain. In order to keep the data transfer
limited and regular, we need to impose a CFL-like condition to restrict the
displacement.
The number of additional halo points needs to be kept small since each additional
halo point requires the exchange of a five-dimensional facet of the six
dimensional hyperrectangle.

\subsection{Fixed-interval Lagrange interpolation}

\begin{figure}
\centering
\setlength{\unitlength}{1.0cm}
\begin{picture}(6.2, 1)
 \put(0, 0.5){\line(1, 0){6.2}}
 \multiput(0.1,0.4)(1,0){7}{\line(0,1){0.25}}
 {\color{magenta}\multiput(1.1,0.5)(1,0){5}{\circle*{0.15}}}
 \put(3.0,0.0){$z_{j}$}
{\color{blue}\put(3.6,0.4){\line(0,1){0.25}}
  \put(3.5,0.0){$z_{j}+\alpha$}}
 \end{picture}
\caption{Fixed-interval Lagrange interpolation based on a five-point stencil.
The red dots indicate the points necessary to calculate the value at $z_j +
\alpha$.}
\label{fig:lagrange_fixed}
\end{figure}
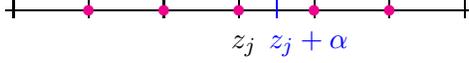

Lagrange interpolation with a stencil that is fixed around the original data point $x_i$ is a very simple interpolation formula for distributed domains. In this case, the interpolation formula with an odd number $q$ of points is given by
\begin{equation}
f(x_j+\alpha) \approx \sum_{i=j-(q-1)/2}^{j+(q-1)/2} \ell_i(\alpha) f(x_i),
\end{equation}
where $\ell_i(\cdot)$ denotes the Lagrange polynomial centered at $i$.  Figure
\ref{fig:lagrange_fixed} illustrates on which data points a five-point stencil
is based.  In this case, we have a static data exchange pattern where
$\frac{q-1}{2}$ points are needed from the processors on the left and on the
right. On the other hand, we need to require $|\alpha|\leq \Delta x$ for
stability, i.e. the scheme is rather restrictive on the time step.

\subsection{Centered Lagrange interpolation}\label{sec:centered_lagrange}

\begin{figure}
\centering
\setlength{\unitlength}{1.0cm}
\begin{picture}(6.2, 1)
 \put(0, 0.5){\line(1, 0){6.2}}
 \multiput(0.1,0.4)(1,0){7}{\line(0,1){0.25}}
 {\color{magenta}\multiput(3.1,0.5)(1,0){4}{\circle*{0.15}}}
 \put(3.0,0.0){$z_{j}$}
 {\color{blue}\put(4.6,0.4){\line(0,1){0.25}}
 \put(4.5,0.0){$z_{j}+\alpha$}}
\end{picture}
\caption{Centered Lagrange interpolation based on a four-point stencil. The red
dots indicate the points necessary to calculate the value at $z_j + \alpha$.}
\label{fig:lagrange_centered}
\end{figure}
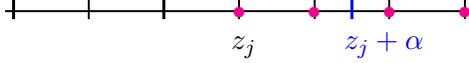

As an alternative, we consider the Lagrange interpolation  for an even number $q$
centered around the displaced point $x_j+\alpha$. Then, the interpolated values
are given as
\begin{equation}
f(x_j+\alpha) = f(x_{j+\gamma}+\beta) = \sum_{i=j+\gamma+q/2-1}^{j+\gamma+q/2} \ell_i(\beta) f(x_i)
\end{equation}
The choice of the data points for the interpolation stencil is illustrated in
Figure \ref{fig:lagrange_centered} for a four-point-formula.  In this case, we
need to exchange a layer of $\max(\frac{q}{2}-\gamma, 0)$ points for the
processor on the left and a layer of $\max(\frac{q}{2}+\gamma-1,0)$  points to
the right. As long as $1-\frac{q}{2} \leq \gamma \leq \frac{q}{2}$,  i.e.
$|\alpha| \leq \frac{q}{2} \Delta x$, the total number of points that need to be
exchanged per stripe is always $q+1$. However, $\gamma$ changes depending on the
value of the other coordinates---and for $\vb$ advections also with time.
Therefore, the communication pattern is different for different stripes. On the
other hand, for Vlasov--Poisson the largest displacements usually appear for the
$\xb$ advections on grid points with high velocities. For an $x_d$ advection,
$d=1,2,3$, the displacement $\alpha = \Delta t v_d$ is very simple and---for
constant time steps---constant in time. In case we require $|\alpha| \leq
\frac{q}{2} \Delta x$ (to retain the minimal communication), the domain can be
split in $q$ blocks of different ranges of $v_d$ with the same interpolation
stencil and, hence, the same data exchange pattern. This way, we can relax the
time step restriction but at the same time keep a regular pattern of both data
exchange and computations for the advections. Note that this is only true for
the case without background magnetic field. The rotation of the grid with the
magnetic field changes the displacement of the $\xb$ advection steps on the
logical mesh.

Having detailed on the mathematical background, we now turn towards a discussion
of the aspects and challenges of an efficient implementation and
parallelization.

\section{Implementation and parallelization}\label{sec:implementation}

Representing the six-dimensional distribution function on a grid consumes a
large amount of memory.  At the same time the arithmetic intensity of a Vlasov
solver is relatively low compared to, for instance, a gyrokinetic solver due to
the less complex structure of the equations.  Therefore, the number of compute
nodes needed for a simulation is determined mostly by the memory requirements.

For our implementation, we use object-oriented Fortran (2003), the Message
Passing Interface (MPI) for distributed-memory parallelism, and OpenMP
directives and runtime functions to add shared-memory parallelism. The
developments were made within the framework of the library
SeLaLib\footnote{Selalib home, \url{http://selalib.gforge.inria.fr/}, accessed on 2018-04-15.}.
The 6d distribution function is discretized on a 6d Fortran array.  Since a
central idea of our method is dimensional splitting, the advection algorithms
exclusively work on one-dimensional stripes of the 6d data.
Since these stripes are non-contiguous in memory for any direction but the first
dimension, non-contiguous 1d slices are copied from the 6d array into a
contiguous buffer before the interpolated values are computed.
The performance of these strided memory accesses can be greatly improved by
cache blocking as is discussed in section \ref{sec:data_access}.

For the domain-decomposition-based parallelization approach, the 1d stripes are
distributed over multiple processes.  The following section discusses how the
halo data is stored in order to perform the interpolations.

\subsection{Distributed-memory parallelism}
\label{sec:parallel_data_layout}

The domain decomposition approach requires a layer of halo cells around the
processor-local data points in order to be able to conveniently compute the
interpolants.  In the discussion below as well as in our Fortran-based
implementation, we consider global indices used locally in each MPI process.

A straight-forward way to handle the halos would be to include the cells into
the 6d array of the distribution function.  Synonymously, this can be regarded
as to work with 6d arrays that overlap between neighboring processors.
Padding the 6d array with the halo cells has the advantage that the data is laid
out contiguously in memory in the first dimension stripe-by-stripe, i.e. the
interpolation routines can work directly on the array in this special case.
Stripes along higher dimensions are conveniently accessed via Fortran-style
linear indexing, however, one has to keep in mind that the elements are laid out
in memory in a strided fashion.  It is important to avoid Fortran array slicing
operators which cause temporary arrays to be used.  Performance can be improved
dramatically by implementing a cache blocking scheme using 2d buffer arrays, see
below.
Moreover, by using halos there is no need for special treatment of periodic
boundary conditions during the advection step.

A second possibility is to allocate the halo cells separately from the 6d array
of the distribution function, i.e. there is no index overlap on the 6d array
between neighboring processes.  Note that in this situation the halo buffers are
identical to the MPI receive buffers.

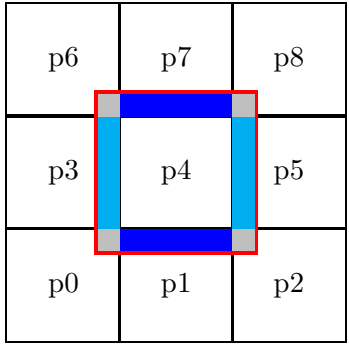
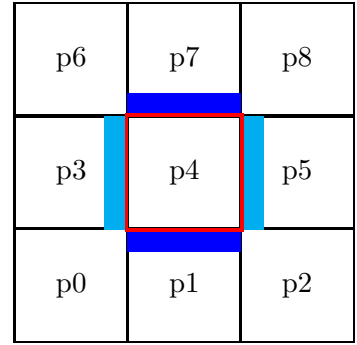
\begin{figure}
	\subfloat[Local arrays with indices that overlap between different processes.]{
	\setlength{\unitlength}{1.5cm}
  \begin{picture}(3,3)(0,0)
          \put(0,0){\framebox(1,1){p0}}
          \put(1,0){\framebox(1,1){p1}}
          \put(2,0){\framebox(1,1){p2}}
          \put(0,1){\framebox(1,1){p3}}
          \put(1,1){\framebox(1,1){p4}}
          \put(2,1){\framebox(1,1){p5}}
          \put(0,2){\framebox(1,1){p6}}
          \put(1,2){\framebox(1,1){p7}}
          \put(2,2){\framebox(1,1){p8}}
          {\put(0.8,1){\setlength{\fboxsep}{0pt}\colorbox{cyan}{\makebox(0.2,1)[r]{}}}
          \put(2,1){\setlength{\fboxsep}{0pt}\colorbox{cyan}{\makebox(0.2,1)[r]{}}}}
          \put(2,1){\setlength{\fboxsep}{0pt}\colorbox{cyan}{\makebox(0.2,1)[r]{}}}
          \put(1,0.8){\setlength{\fboxsep}{0pt}\colorbox{blue}{\makebox(1,0.2)[r]{}}}
          \put(1,2){\setlength{\fboxsep}{0pt}\colorbox{blue}{\makebox(1,0.2)[r]{}}}
          \put(0.8,2){\setlength{\fboxsep}{0pt}\colorbox{lightgray}{\makebox(0.2,0.2)[r]{}}}
          \put(0.8,0.8){\setlength{\fboxsep}{0pt}\colorbox{lightgray}{\makebox(0.2,0.2)[r]{}}}
          \put(2,2){\setlength{\fboxsep}{0pt}\colorbox{lightgray}{\makebox(0.2,0.2)[r]{}}}
          \put(2,0.8){\setlength{\fboxsep}{0pt}\colorbox{lightgray}{\makebox(0.2,0.2)[r]{}}}
          {\linethickness{1pt}{\color{red}\put(0.8,0.8){\framebox(1.4,1.4){}}}}
  \end{picture}}
  \hfill
  \subfloat[Local arrays with indices that do not overlap between different processes.]{
	\setlength{\unitlength}{1.5cm}
  \begin{picture}(3,3)(0,0)
          \put(0,0){\framebox(1,1){p0}}
          \put(1,0){\framebox(1,1){p1}}
          \put(2,0){\framebox(1,1){p2}}
          \put(0,1){\framebox(1,1){p3}}
          \put(1,1){\framebox(1,1){p4}}
          \put(2,1){\framebox(1,1){p5}}
          \put(0,2){\framebox(1,1){p6}}
          \put(1,2){\framebox(1,1){p7}}
          \put(2,2){\framebox(1,1){p8}}
          \put(2,1){\setlength{\fboxsep}{0pt}\colorbox{blue}{\makebox(0.2,1)[r]{}}}
          {\put(0.8,1){\setlength{\fboxsep}{0pt}\colorbox{cyan}{\makebox(0.2,1)[r]{}}}
          \put(2,1){\setlength{\fboxsep}{0pt}\colorbox{cyan}{\makebox(0.2,1)[r]{}}}}
          \put(1,0.8){\setlength{\fboxsep}{0pt}\colorbox{blue}{\makebox(1,0.2)[r]{}}}
          \put(1,2){\setlength{\fboxsep}{0pt}\colorbox{blue}{\makebox(1,0.2)[r]{}}}
          {\linethickness{1pt}\color{red}\put(1,1){\framebox(1.0,1.0){}}}
  \end{picture}}
\caption{Schematic diagram of a data layout example, simplified to two
dimensions and distributed over 9 MPI processes labeled p0 \ldots p8. The square
tiles represent the processor-local parts, the blue and cyan blocks the halo
data needed for the advections along the first and the second dimension.  The
red square shows the data array stored by processor p4 for the two different
data layouts.  Layout (a) allocates the halo cells as part of the data array.
Layout (b) allocates the halos as independent arrays.  For layout (a), the gray
blocks in the corners are unused.}
\label{fig:data_layout}
\end{figure}
Figure \ref{fig:data_layout} shows these two possibilities.
In a 6d array which includes the halo cells, one also has to allocate the corner
points (displayed gray in Figure \ref{fig:data_layout}(a)) as well, even though
they are not used by any of the 1d interpolators.  As is well known, the volume
of a hypercube mostly concentrates in the corners, therefore it is desirable to
avoid memory allocation there.
Moreover, in case the algorithm uses a different number of halo cells on
different blocks of data, the 6d array has to include the maximum number of halo
cells in any direction.
Therefore we have chosen the second approach, using halo cells allocated
separately from the 6d array of the distribution function. It avoids the
aforementioned disadvantages and is in particular superior with respect to
memory efficiency.
Moreover, we can also exploit the fact that we only need the halo cells along
one dimension at a time.  Once allocated the halo buffers can be reused.

If we assume a MPI-process-local grid of size $N^6$ and a halo width of $w$
cells on each side of the domain, the basic memory requirements for the domain
decomposition scheme are $N^6 + 2 w N^5$ (for the core 6d array and two halo
buffers of size $w N^5$ on each side).  Note that an additional send buffer of
the size of a single halo buffer is needed.
In a neighbor-to-neighbor communication, two data blocks of size $w N^5$ need to
be communicated for each one-dimensional advection step.
For the remap scheme, on the other hand, two copies of the local $N^6$ buffer
are needed and, in addition, MPI send and receive buffers. Between each block of
$\xb$ and $\vb$ advections, data blocks of size $\frac{1}{p}N^6$ have to be
communicated to each of the $p-1$ other MPI processes, i.e. a total fraction of
$\frac{p-1}{p}$ of the local data block is sent.
In practice the actual memory requirements may be even higher due to
MPI-internal buffers.

\begin{table}[htbp]
\caption{Comparison of the theoretical minimum memory requirements per MPI process
for both algorithms under consideration.  A distribution function at a
resolution of $N^6$ local points is considered, in addition we assume for the
domain decomposition case (d.d.) two halo buffers with a width $w=3$ points each.  Note
that additional buffers which may be required e.g. by the MPI library, are not
accounted for.  Moreover, the table shows for each resolution the data volume
 that is communicated per time step per process ($p\rightarrow \infty$ for remap).}
\label{tab:memory_theoretical}
\begin{tabularx}{\columnwidth}{l|ll|ll}
\hline
\; & \multicolumn{2}{|l|}{allocated mem. [GiB]} &
\multicolumn{2}{|l}{communicated mem.
[GiB]} \\ $N$ & remap & d.d. & remap & d.d. \\
\hline
\hline
16 & 0.25 & 0.17 & 0.5 & 0.28 \\
32 & 16.00 & 9.50 & 32.00 & 9.00 \\
40 & 61.04 & 35.10 & 122.08 & 27.47 \\
64 & 1024.00 & 560.00 & 2048.00 & 288.00 \\
\hline
\end{tabularx}
\end{table}

Table \ref{tab:memory_theoretical} compares the theoretical memory requirements
for a typical process-local number of points per dimension for the two memory
layouts. Note that the remap parallelization uses two 6d data arrays for the
two remap data layouts.  The comparably low memory consumption of the domain
decomposition implementation is especially advantageous on systems where fast
memory is a scarce resource, e.g. on certain manycore chips.

Moreover, based on the numbers from table \ref{tab:memory_theoretical} one can
give a straightforward estimate of the resolution possible on a cutting-edge 
HPC system with $\sim 100$ GB memory per two-socket node.
Considering the domain decomposition algorithm and putting two MPI processes per
node with $32^6 - 40^6$ points each, a grid size of $128^6 - 160^6$ would fit
on 2048 nodes. Further increasing the resolution, e.g., in velocity space,
problems of size $128^3 \cdot 256^3 - 160^3 \cdot 320^3$ would fit on 16384
nodes.





\subsection{Data access for 1d interpolations}\label{sec:data_access}

On the distributed domain, the advection along a dimension takes the
form shown in Algorithm \ref{alg:advection_pure} at the example of $x_3$.
Note that for advections along the dimensions 2 to 6 we have to deal with
increasingly large strides when the 1d interpolation buffers are filled, causing
a severe performance penalty due to cache misses as confirmed by profiling.

The cache efficiency and the resulting performance can be greatly improved by a
cache-blocking scheme similar to the schemes used to accelerate, e.g., dense
linear algebra operations.
The blocking is based on a 2d buffer array.  Interpolations are performed along
the first (contiguous) dimension.  In the second dimension the array is large
enough to store at least a cache line of data.
Algorithm \ref{alg:advection_blocked} summarizes the loop rearrangements. A
similar blocking has been implemented for all advection steps. For the
advections along $x_2$ to $x_6$, we extract the one-dimensional stripes in
blocks along the first dimension.

\begin{algorithm}[htbp] Copy strided data to contiguous send buffer\; MPI communication of halos\;
\For{i6}{ \For{i5}{ \For{i4}{ \For{i2}{ \For{i1}{ Copy 1d stripe over i3 from 6d
array and halo buffers into buffer\; Interpolation along x3\; Copy 1d stripe
with updated values back to 6d array\; } } } } } \caption{Advection along $x_3$
without cache blocking.} \label{alg:advection_pure} \end{algorithm}

\begin{algorithm}[htbp] \caption{Advection along $x_3$ with cache blocking.}
\label{alg:advection_blocked}
Copy strided data to contiguous send buffer\; MPI communication of halos\;
\For{i6}{ \For{i5}{ \For{i4}{ \For{i2}{ \For{i3}{ \For{i1}{ Copy data from 6d
array and halo buffer into 2d buffer\; } } \For{i1}{ Interpolation along x3\; }
\For{i3}{ \For{i1}{ Copy interpolated data from 2d buffer back to 6d array\; } }
} } } } \end{algorithm}

\subsection{Shared memory parallelization}

Both implementations, remap and domain partitioning, are carefully parallelized
using OpenMP directives and runtime functions to exploit the shared-memory
architecture of prevalent multicore CPUs using threads, in addition to the
distributed-memory parallelization which employs MPI processes.

A significant advantage of introducing a hybrid parallelization in addition to
MPI is that it allows to reduce the memory consumption and the communication
volume significantly.
Instead of running one MPI process per available processor core, each allocating
and exchanging halo cells, it is superior to launch only one or two MPI
processes per socket, each with a proportionate number of threads pinned to the
cores of that socket.
All threads thereby share the halo cells.

Let us illustrate this effect by giving a simple numerical example.  Consider a
$64^6$ simulation using 7 point Lagrange interpolation that shall be run on 64
compute nodes with 64 cores each, implying a grid size of $32^6$ points per
node.  If a plain MPI setup is chosen, each node would run 64 MPI processes with
a local grid size of $16^6$ points, totaling up to 4096 MPI processes.  On the
other hand, we might consider an (extreme) hybrid setup running only 1 MPI
process per node with a local grid size of $32^6$ points.  The $64$ processes of
the plain MPI setup would allocate 11 GB of memory per node for the distribution
function and the halo cells, and communicate 36 GB per time step, 18 GB of which
beyond the node over the network.  In contrast, the hybrid setup would require
9.5 GB per node and communicate 18 GB over the network.

We conclude that, first, it may be advantageous to use as few MPI processes as
possible from the memory and communication point of view.  Second, while the
hybrid setup eliminates intra-node communication the inter-node communication
volume stays the same compared to the plain MPI case, with larger message sizes
though.

A potential disadvantage of a naive hybrid approach is due to the fact that a
significant fraction of the threads would be idle during the data-intense halo
exchanges, however, by introducing an advanced pipelining scheme we are able to
hide the communication times to a large extent as is discussed in the following
section.

\section{Performance optimization}\label{sec:optimizations}

Performance optimization work aims at maximizing the node-level performance
simultaneously with the parallel scalability which are conflicting goals to some
degree.
In the scope of this paper, we target recent x86\_64 systems with multi-core or
many-core CPUs as found in the vast majority of today's HPC systems.  Details on
the systems under closer consideration are given in Table \ref{tab:nodes} in the
following section where performance results will be discussed.


\subsection{Performance profile}

\begin{figure}
\includegraphics[width=\columnwidth]{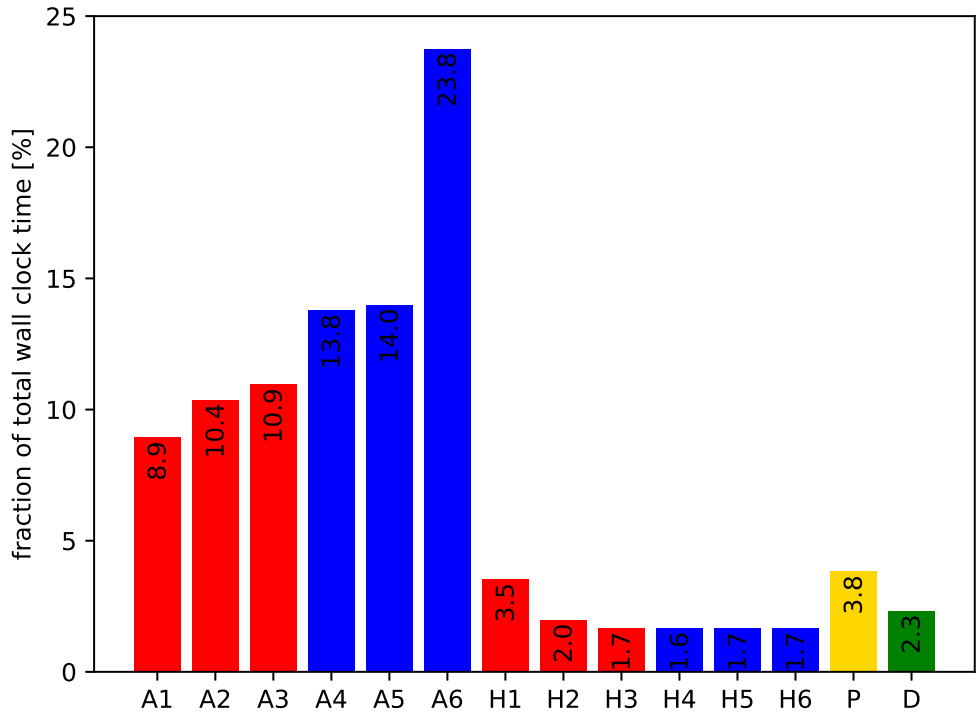}
\caption{Profile of the domain decomposition implementation running a $32^6$
case with 7pt Lagrange interpolation on a single Haswell core, where the letter
``A'' denotes advection, the letter ``H'' denotes a halo-exchange operation, the
letter ``P'' denotes the Poisson solve operation, the letter ``D'' denotes the
diagnostic computation, and the direction is given by its number.}
\label{fig:single_core_clocks}
\end{figure}

Fig.~\ref{fig:single_core_clocks} shows a breakdown of the costs of the various
operations involved during a time step of the domain decomposition solver
running on a single core without any parallelism.  The profile is
clearly dominated by the advection computations (``A'') including the Lagrange
interpolation.  Going from the direction of the first to that of the sixth
dimension, the cost of the advection monotonically increases.  This effect is caused by
the fact that memory accesses become more and more strided. It is important to
note that the effects of the striding are already mitigated by the cache
blocking scheme which preserves a cache line, once loaded.  Moreover,
the prefetch efficiency of the processor appears to deteriorate with increasing
strides.  The preparation of the halo buffers (``H''), which also involves copies from
strided into contiguous memory (however of much less data compared to step ``A''), is by far less time-consuming. However,
neighbor-to-neighbor MPI communication is included in ``H'' which becomes
important when the parallelization spans multiple nodes. Finally, the
Poisson-solve step and the diagnostics account for roughly 4 and 2 percent
of the total runtime, respectively.

In addition to simple and lightweight timing facilities, we used the tools
Amplifier and Advisor from the Intel Parallel Studio XE package to obtain low
level information on the performance and limitations of various parts of the
code.  Based on that information, code improvements were implemented, the most
important of which are detailed on below.

\subsection{Single-core performance}

Without considering MPI communication, the major factors limiting performance of
both the 6d Vlasov implementations, the remap and the domain decomposition, are
due to the fact that the vast majority of the memory accesses---except the
ones along the first dimension---are strided.
A cache blocking scheme mitigates this issue significantly, as illustrated by
performance numbers below.  Nevertheless the code remains memory bound.  The
increase in run time from ``A1'' to ``A6'' in Fig.~\ref{fig:single_core_clocks}
reflects the aspect of the increasingly strided memory accesses.

In addition, SIMD vectorization is a key factor to achieve performance on modern
CPUs.  While in early days (SSE2) only a factor of two was lost when
vectorization was ignored for double-precision operations, the potential loss
has grown to a factor of 4 (AVX) or 8
(AVX512) on more recent CPU models.
We have implemented Lagrange interpolation routines such that the compiler is
able to generate vectorized code which we verified carefully using compiler
reports and performance tools.  Arrays are aligned to 64 byte boundaries, though
the large 6d array of the distribution function is not padded in order not to
waste memory.  In general, the compiler is able to generate vectorized code for
most of the loops.


Running 100 time steps of a $16^6$ ($32^6$) case with 7pt Lagrange interpolation
on a single Skylake core, the domain decomposition code achieves a floating point operation rate of
$3.9$ ($2.8$) GFLOP/s, which translates to $2.5$ ($2.5$) GFLOP/s on the Haswell
core.\footnote{Note that the smaller setup is relatively faster on the Skylake
CPU. We measured the FLOP rates using performance counters on Skylake and
converted the result to Haswell using the runtime ratio.} This value represents
about $6.8\%$ of the Haswell core's theoretical peak performance.
Note that the measurement covers the complete run including startup and shutdown
phases and includes inevitable memcopy operations that do not perform any FLOPs
at all.  Around 90\% of the floating point instructions issued are vectorized.
These numbers once more illustrate the main performance challenge of 6d Vlasov
codes resulting from memory boundedness due to strided memory accesses in
combination with a moderate arithmetic intensity.

Finally, to quantify the effect of the cache blocking algorithm, the
aforementioned test runs with 100 time steps take on the Haswell core in total
about a factor of $2.4$ longer for both cases with the cache blocking disabled.
The higher the dimension to be interpolated along, the more effective and
important the cache blocking becomes in general, accelerating certain parts of
the code such as the loop over $x_6$ by up to a factor of $20$, as measured
using performance profilers.

\subsection{Node-level performance}

A modern compute node provides several cores that are organized in non-uniform
memory access (NUMA) domains such that groups of cores share L3 caches and
memory channels these cores can access fastest.  Optimizing for the NUMA domains
by careful process and thread pinning at runtime turns out to be important.
Typically, MPI processes are pinned to sets of cores on sockets, and threads are
pinned to individual physical cores from these sets.  When overlapping
communication and computation as introduced in the following subsection, it
turns out to be advantageous only to pin the processes to constrain the threads
within NUMA domains, and in addition, to use hyperthreads.

As the typical structure of the code are 6-fold nested loops, a standard
loop-based OpenMP parallelization strategy proved rather successful.
From benchmark measurements at various resolutions, it was concluded that the
runtime is minimized when the outer two loops are collapsed into a single one to
increase the granularity of the parallel workload, and when static loop
iteration scheduling is used.
Each thread is able to benefit from cache blocking and vectorization in the
inner loops. Virtually any workload in the code is parallelized using that
technique.  As a result, the application scales well over a single node in pure
OpenMP as shown in Sec.~\ref{sec:performance}.

\subsection{Distributed-memory performance}\label{sec:pipelining}

Semi-Lagrangian 6d Vlasov solvers are unavoidably intense in terms of memory and
data communication volume (cf.~Sec.~\ref{sec:parallel_data_layout}).  For the
domain decomposition approach, typical sizes of single MPI messages are in the
range of $\mathcal{O}(0.1)$--$\mathcal{O}(1)$ GB, while modern interconnects
achieve a bandwidth of up to $\approx 10$ GB/s per node.  It is therefore
important to mitigate the cost of the data transfers as much as possible,
firstly by careful planning of the process grid, by overlapping communication
and computation, and in addition by means of data reduction.

As outlined in Algorithm \ref{alg:advection_blocked}, each advection step starts
with MPI communication in order to fill the halo buffers in the neighbor
processes, before the interpolations are performed.  In hybrid-parallel setups
the initial MPI communication would only keep a single thread busy while all the
other threads were idle.  The trend towards systems with increasingly more cores
per socket suggests to use multiple threads per MPI process in order to overlap
(``pipeline'') the communication with useful computation.

\subsubsection{Simultaneous communication and computation.}

\begin{figure}
\includegraphics[width=\columnwidth]{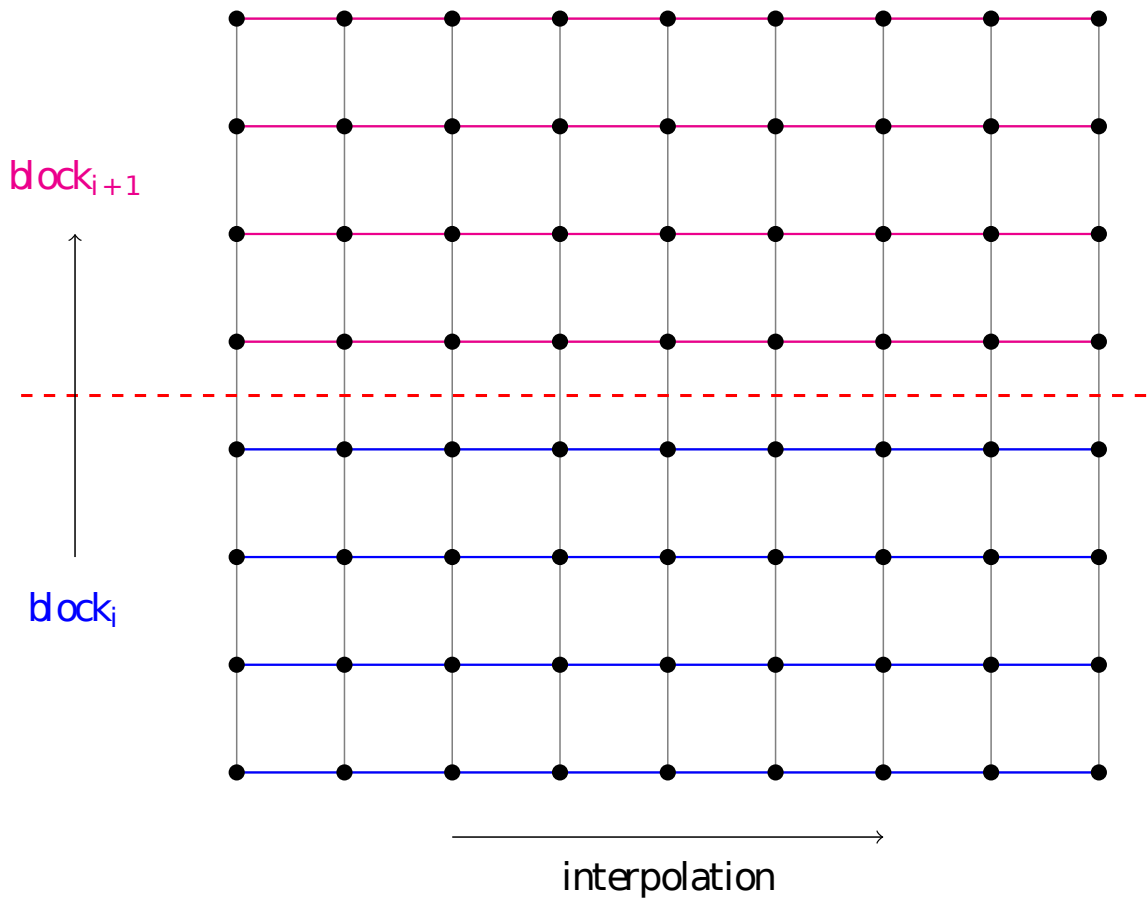}
\caption{Blocking of a 2d example grid, perpendicular to the direction along
which is to be interpolated.}
\label{fig:blocking}
\end{figure}

In order to implement pipelining of communication and computation, we block the
data along a dimension different from the one we intend to interpolate along,
and perform data exchange and computation simultaneously on the resulting
independent blocks, as illustrated in Fig.~\ref{fig:blocking}.

Here, we consider the Lagrange interpolation with fixed interval because in this
case we do not have to handle additional blocking due to asymmetric data
exchange.  Moreover, in order to avoid a second layer of blocking, we do not
consider blocks with different halo patterns.  Anyway, the overhead introduced
by not minimizing the halo widths for some blocks is less problematic when the
communication is overlapped with computations.

For each data block, the advection computation consists of three steps:
Copy (generally non-contiguous 6d) data into coherent buffer (C);
MPI\_sendrecv() communication (M); computation of interpolated values
(I).

\begin{figure}
\includegraphics[width=\columnwidth]{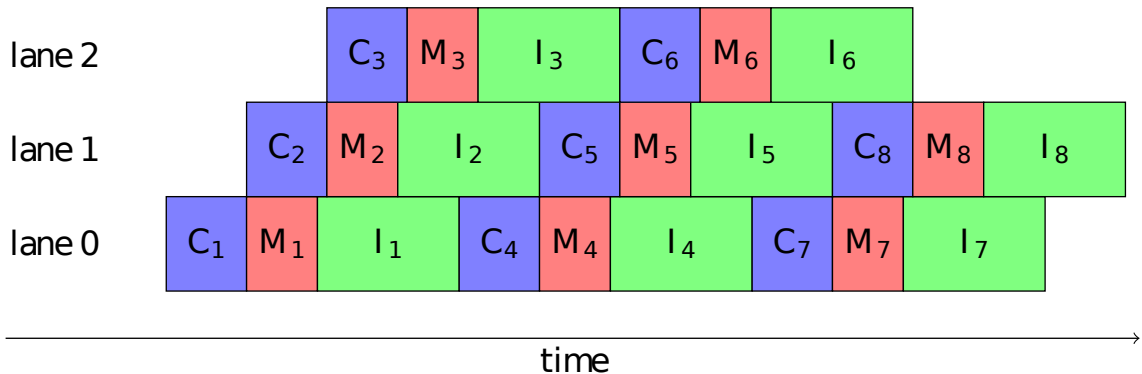}
\caption{Schematic showing the temporal overlap of the copy operation (C), the
MPI communication (M), and the interpolation (I) for a single MPI process at the
example of an 8-fold blocked advection.}
\label{fig:task_graph}
\end{figure}

For each data block, these three steps need to be performed in the given order,
but there is no dependency between different blocks. Nevertheless, we have to
enforce some ordering in order to avoid a capacity overload of resources such as
the maximum number of simultaneous hardware threads.  Given the 3-fold structure
of the advection, we propose a straight forward pipelining scheme using 3 lanes
as shown in Fig.~\ref{fig:task_graph}.

In order to keep the overhead of the start-up and the final phase as small as
possible, we overlap communication and computation of advections for different
dimensions as well. To give an example let us start with the $x_1$ advection.
Once we have reached the communication step (``M'') of its last block, we can
already start to copy (``C'') and communicate (``M'') the data needed for the
following $x_2$ advection in the first data block---provided that the $x_2$
dimension is contiguous in each block.

\subsubsection{Implementation details on the pipelining scheme.}

Our pipelining implementation relies on a thread-safe MPI library and on
OpenMP threads---requirements that are provided by most modern compilers and
libraries.  The steps ``C'', ``M'', and ``I'' can be regarded as tasks
with interdependencies.

\begin{algorithm}[htb] Build list of N blocks, add dummy block, initialize
OpenMP locks\; Enter OpenMP parallel region with 3 threads\; \For{i (OpenMP:
schedule static, chunk size 1) $\leftarrow$ 1 \KwTo N} { Set lock of C(i+1)\;
Set lock of M(i+1)\; } \For{i (OpenMP: schedule static, chunk size 1)
$\leftarrow$ 1 \KwTo N} { Set lock of C(i)\; C$\dagger$(i)\; Unset lock of
C(i)\; Unset lock of C(i+1)\; Set lock of M(i)\; M(i)\; Unset lock of M(i)\;
Unset lock of M(i+1)\; I$\dagger$(i)\; } Leave OpenMP parallel region\; Destroy
locks\; \caption{Pipelining algorithm as implemented using OpenMP directives and
runtime functions. The symbol $\dagger$ indicates the use of nested
parallelism.} \label{alg:overlap} \end{algorithm}
In the following, we provide details on the implementation, referring to
Algorithm \ref{alg:overlap}.  Initially, a list of all blocks is built, where
each list element contains meta data such as block indices, the number of nested
threads for the steps ``C'' and ``I'', and two OpenMP locks, one for the ``C''
step and one for the ``M'' step.  In the scheme proposed in the following, the
orchestration of the tasks is explicitly controlled using a parallel region with
a fixed number of three threads which uses internal loops with static scheduling
and a chunk size of 1 such that the ordering is deterministic.
For $N$ blocks, thread $i$, with $i=0,1,2$, manages the work on the blocks
$\ell=i+1+3k$, $k=0,\ldots, \lfloor\frac{N-i-1}{3}\rfloor$, performing the steps ``C'', ``M'',
and ``I'' one after the other. In the steps ``C and ``M'', we use nested OpenMP
parallelism to make use of all the available threads. Apart from managing locks
and calling the tasks the outer loop does nothing. In order
to make best use of the network resources, we ensure that the resources are
dedicated to one ``C'' and one ``M'' operation at a time, i.e.~only a single
communication ``M'' is allowed to take place at a time.
Therefore, thread $i$ initially locks all the ``C'' and ``M'' operations of thread
mod($i+1$,3) and only releases the lock on ``M$_{\ell+1}$''/``C$_{\ell+1}$'', when it has finished
``M$_{\ell}$''/``C$_{\ell}$''.
%
As a result, the pipelining scheme with overlapping tasks as shown in
Fig.~\ref{fig:task_graph} arises.

At runtime, we have to specify two parameters that
influence the performance of the implementation: the number of blocks and the
number of threads used by each of the ``C'' and ``I'' tasks. The smaller the blocks
are chosen the shorter the start-up and finishing phases become during which
communication and computation cannot be overlapped.  On the other hand, the
blocks should not be chosen too small in order to keep the overhead low.  As the second parameter, one needs to decide how to
assign the available threads to the ``C'' and ``I'' steps. Note that the first ``C''
and the last ``I'' step may use more threads since they do not fully overlap with
other operations.  It turns out that the use of simultaneous multithreading
(``hyperthreading'') in concert with the pipelining is beneficial.  A detailed
discussion including benchmark results is presented in Section
\ref{sec:performance}.

An obvious choice for the advections in $\xb$ is to block the 6d distribution
function along the 6th dimension $v_3$ which corresponds to the slowest varying
loop index resulting in blocks that are laid out contiguously in memory.  Thus
our pipelining scheme combines the three $\xb$ advection steps without any inner
global synchronization point.  In the same way, the $\vb$ advection block can be
combined. Here the blocking of the data is performed along $x_3$.


\subsubsection{Optimization of the halo-exchange communication.}

As pointed out before, the 6d Vlasov solver is a highly communication intensive
application, rendering good large-scale parallel scalability a difficult goal to
achieve. From our experiments, we find that the implementation can scale well
even when spanning multiple islands on a HPC system if the process grid is laid
out in an optimum way and communication and computation are pipelined.

The HPC systems under consideration have the network topology of a pruned fat
tree.  An island of the system is designed such that each node of the left half
of the island is theoretically able to communicate with the corresponding node
of the right half of the island simultaneously at the same bandwidth, termed the
bisectional bandwidth.
%
Going beyond an island, the bisectional bandwidth is reduced.
Here, the blocking factor is decisive, which is, e.g., 1:4 on a system used
below, meaning that the inter-island bisectional bandwidth is only a quarter of
the intra-island one.
We are therefore challenged with at least four levels of interconnection speeds
between MPI processes, namely communication via shared memory on the same
socket, intra-node communication via shared memory between different sockets,
intra-island and inter-island communication via the network.

In addition to the pruned fat tree topology, there are other topologies, e.g.,
high dimensional torus interconnects.  We would expect the 6d domain
decomposition algorithm to benefit highly from such networks because a reduced
bisectional bandwidth is avoided, whereas on a pruned fat tree it turns out to
be one of the major challenges to scalability.

It is crucial to optimize the Vlasov solver and also choose the problem setup for
the network topology of the machine as well as possible in order to prefer fast
communication paths for the largest messages, which is discussed in the
following.

\subsubsection{Process grid optimization.}

The ordering of the 6d process grid can be chosen such that communication
between remote processes is kept as small as possible. A batch system lays out
the MPI processes in a certain pattern, often placing the ranks consecutively on
the nodes, one island after the other. When constructing the 6d process grid,
the MPI ranks are placed in row-major (C) order, indexing processes $p$ like
$p[i,j,k,l,m,n]$, with $n$ being the fastest varying index.  Consequently, the
neighboring processes are closest in $x_6$, direction $n$, whereas they are
separated increasingly on the network when going to the $x_1$, direction $i$.
Depending on the shape of the process-local part of the 6d grid and on the
(potentially different) interpolation orders along $\xb$ and $\vb$, the process
grid layout can be chosen such that inter-island communication is minimized.
Despite this optimization, certain directions still communicate mainly between
islands. In our experiments, it turned out to be beneficial to
transpose the grid such that neighboring processes are closest in $x_1$. In
this case, the processor groups that solve the Poisson problem are close. On the
other hand, the reductions over velocity to compute $\rho$ combine more remote
processes.  When overlapping communication and computations this ordering is
especially advantageous since the communication is more expensive the larger the
stripe, i.e.~the longer it takes to gather the data for the advection.
%

To go one step further, we have experimented with communication patterns between
islands that are more balanced in the time dimension. This is done by mapping
consecutive MPI ranks onto blocks of $2^6$ processes, i.e. by rearranging the 6d
process grid completely.  Our implementation uses known hostname schemes of HPC
systems to perform the rearrangement.
As the consequence of such rearrangement the advection in each direction would
perform inter-island communication to some fraction, which would be useful to
mitigate situations without rearrangement when only one or few directions are
communicated between islands not hiding well behind computation.
%
However, within the scope of this paper we did not enable such blocking because
it did not turn out beneficial, the reason being that the per-process
computational workload (the local partition of the 6d hypercube) was typically
too small in relation to the halos.

It is important to point out that these process grid optimizations do not reduce
the total amount of data that is to be communicated. We will turn towards
possibilities to reduce the communicated data volume in the following section.

\subsubsection{Reduction of the communication volume.}

The MPI messages sent between neighbor processes to fill the halos use by
default 8 byte-wide double precision numbers. We have implemented the option to
halve the communication volume and time by sending these messages in single
precision, leading to an improved parallel scalability especially in
inter-island scenarios. However, one has to keep in mind that an additional
error is introduced into the computation which requires careful validation. We
therefore consider single precision messages an experimental tool.

When running the pipelined code in a hybrid fashion it turns out
that---depending on the balance between threads and processes---a fraction of
the threads is idle waiting for communication to finish.  To continue along the
lines of single precision messages, we have therefore experimented with floating
point compression in order to further reduce the communication volume and time
while utilizing the threads that would otherwise be idle.  Naturally, a simple stream
compression algorithm is not suitable to compress floating point data, rather
a lossy algorithm tailored towards numerical data is required.  The ZFP
floating point compression algorithm, of which an implementation is freely
available as a library, compresses blocks of 64 double precision numbers,
taking the desired precision as a parameter \citep{ZFP}.  The compression ratio
achieved depends on the similarity of the numbers in the blocks.  Adjusting
the precision to match single precision, we typically observe a compression
factor of $\sim 4$ for the halo data, improving parallel scalability over
islands significantly.
%
In particular the compression step is rather expensive as will be shown briefly
in the following.  The compression option might become more relevant in the
future when the per-node computational power continues to grow faster than the
network bandwidth.

In section \ref{sec:performance}, we present performance studies, detailing on
the various challenges and the respective optimization approaches to tackle
them.

\section{Numerical experiments}
\label{sec:numerics}

In this section, we consider representative benchmark problems with and without a background magnetic field and discuss accuracy and CFL-like conditions for the various Lagrange interpolations on distributed grids.

\subsection{Vlasov--Poisson simulations}

We first consider two test problems with no magnetic field, weak Landau damping, where the chosen resolution yields rather good accuracy, and a bump-on-tail instability with larger phase-space error, and compare the accuracy of the Lagrange interpolators of various order.

The initial value of the weak Landau damping problem is given as
\begin{equation}
\begin{aligned}
f_0(\xb,\vb) = \frac{1}{(2\pi)^{3/2}} & \exp\left(-\frac{|\vb|^2}{2}\right) \\
                                      & \left(1+ 0.01\sum_{\ell=1}^3 \cos(0.5x_{\ell})\right).
\end{aligned}
\end{equation}
The grid resolution is chosen to be $16^3 \times 64^3$ and we consider the error in the electrostatic potential energy in the time interval  $[0,30]$ compared to a reference simulation on a grid with  $20^3 \times 80^3$ points and a time step of $\Delta t = 0.005$ with Lagrange interpolation of order 8 in space and 7 in velocity.

The bump-on-tail test case has the initial value
\begin{equation}
\begin{aligned}
& f_0(\xb,\vb) = \frac{1}{(2\pi)^{3/2}} \\
& \left( 0.9 \exp\left(-\frac{v_1^2}{2}\right) + 0.2 \exp\left(-2(v_1-4.5)^2 \right)\right) \\
& \exp\left(-\frac{(v_2^2+v_3^2)^2}{2}\right)\left(1+ 0.03 \sum_{\ell=1}^3 \cos(0.3x_{\ell})\right).
\end{aligned}
\end{equation}
and is solved on a mesh with 32 points per direction until time 15. The reference solution is simulated on a mesh with 40 points per direction, a time step of $\Delta t=0.0125$, and Lagrange interpolation of order 8 in space and 7 in velocity.

\begin{figure}
\includegraphics[width=\columnwidth]{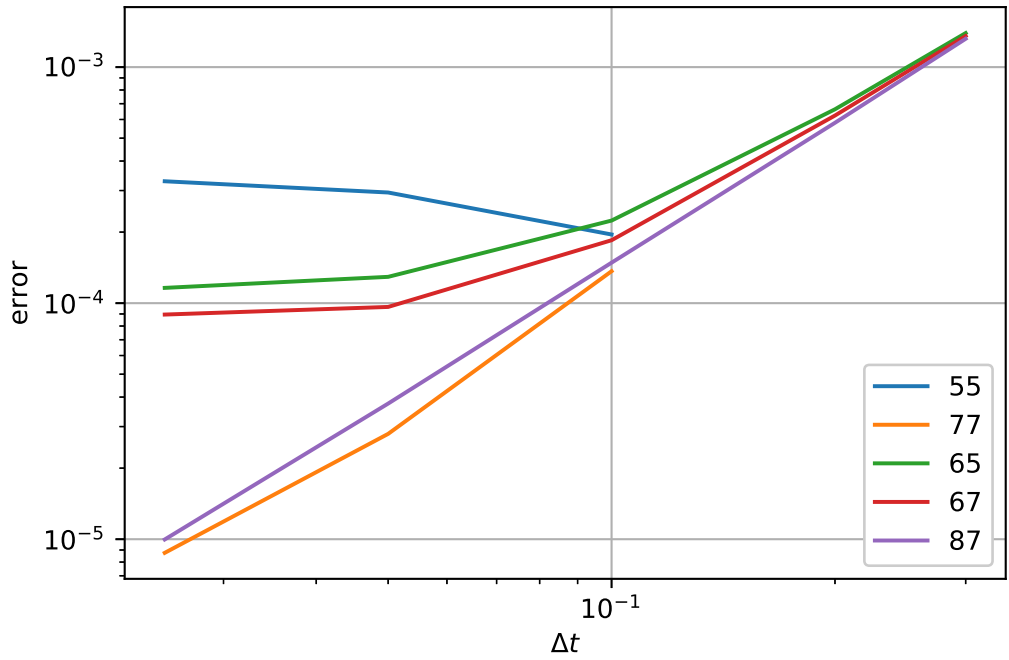}
\caption{Landau damping: Error in electrostatic energy as a function of the time
step for various interpolators.  The numbers indicate the stencil widths of
the $\xb$ and the $\vb$ interpolations, respectively.}
\label{fig:landausum_dt_error}
\end{figure}

Figure \ref{fig:landausum_dt_error} shows the error in the electrostatic energy as a function of the time step for various orders of the Lagrange interpolator for the weak Landau damping example. Comparing the curves, we see that the error for the largest time step $\Delta t = 0.3$ is almost the same for all considered interpolation formulas, that is the temporal error dominates. Since we use a second order Strang splitting method, the error reduces proportional to $\Delta t^2$ as we reduce the time step until the interpolation error starts to dominate. The lower the order of the interpolation stencil the earlier this happens. Note that, once we have reached a time step where interpolation errors dominate, a further reduction of the time step may even yield an increase of the error.

For this example, the displacement of the $\xb$ advection exceeds one cell size for a time step above $0.13$. Therefore, fixed Lagrange interpolation (odd order) is only stable for time steps smaller than 0.13. On the other hand, centered Lagrange interpolation shows good results for time steps beyond this. The results also show that for accuracy reasons the time step should be chosen such that the displacement is on the order of the cell size or a bit above. Hence,  a combination of centered Lagrange interpolation with blocked communication as described in Section \ref{sec:centered_lagrange} for the $\xb$ advections and fixed Lagrange interpolation for the $\vb$ advection is an efficient choice for the Vlasov--Poisson equation.

\begin{figure}
\includegraphics[width=\columnwidth]{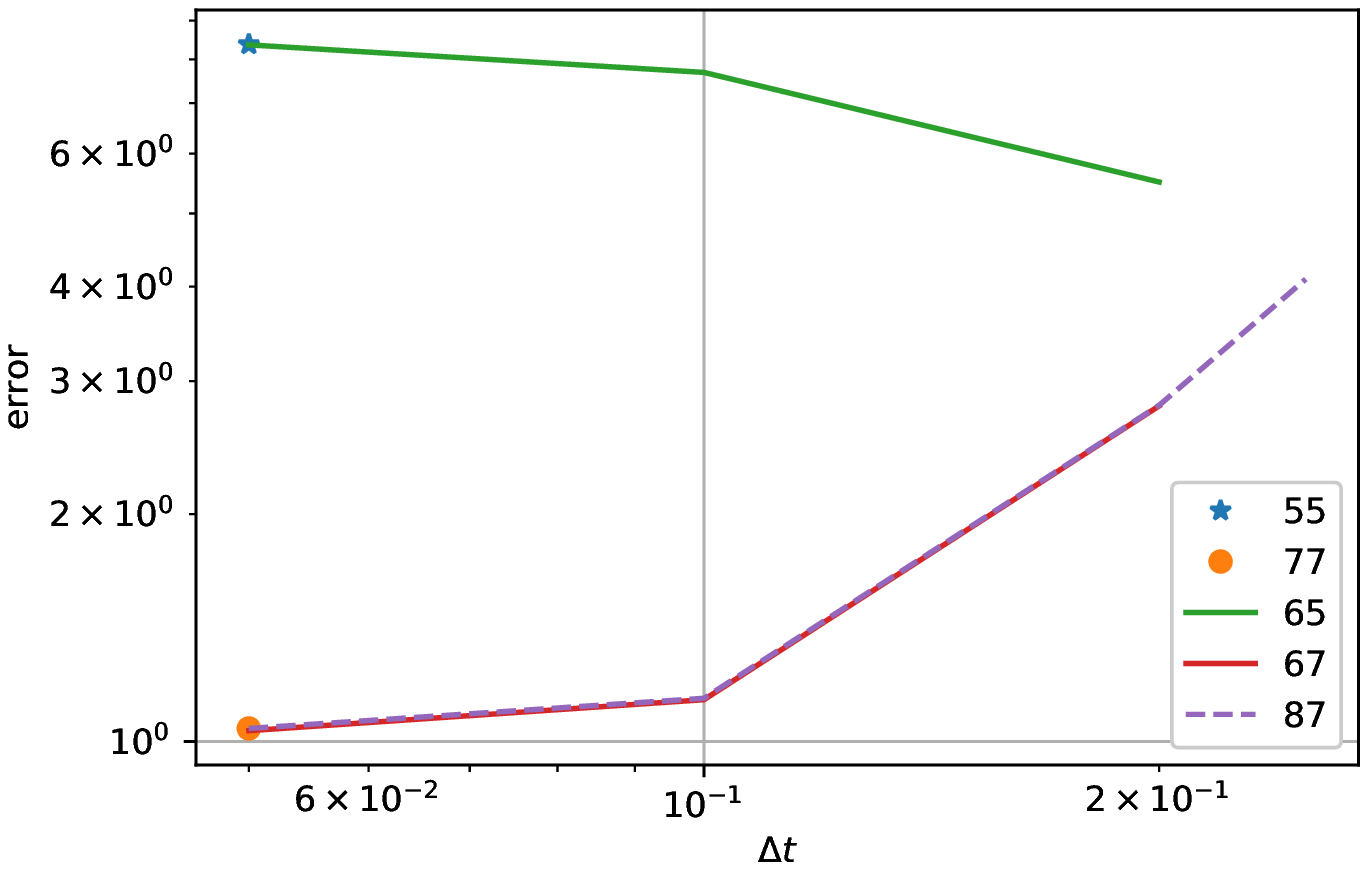}
\caption{Bump-on-tail: Error in electrostatic energy as a function of the time step for various interpolators.  The numbers indicate the stencil widths of
the $\xb$ and the $\vb$ interpolations, respectively.}
\label{fig:bot_dt_error}
\end{figure}

Figure \ref{fig:bot_dt_error} shows the results for the bump-on-tail test case. In this case, the spatial error is larger which is why larger time steps compared to the value of $\Delta t \approx 0.073$ (which corresponds to the maximum time step where the displacement is bounded by one cell size) are advantageous. As in the previous case, we see that order 5 does not give satisfactory results and the use of Lagrange interpolation of order 6 in $\xb$ and 7 in $\vb$ gives best results. Note also that the absolute error is shown in the figure and the maximum value of the electric field energy in the considered interval $[0,15]$ is about 205.

\begin{figure}
\includegraphics[width=\columnwidth]{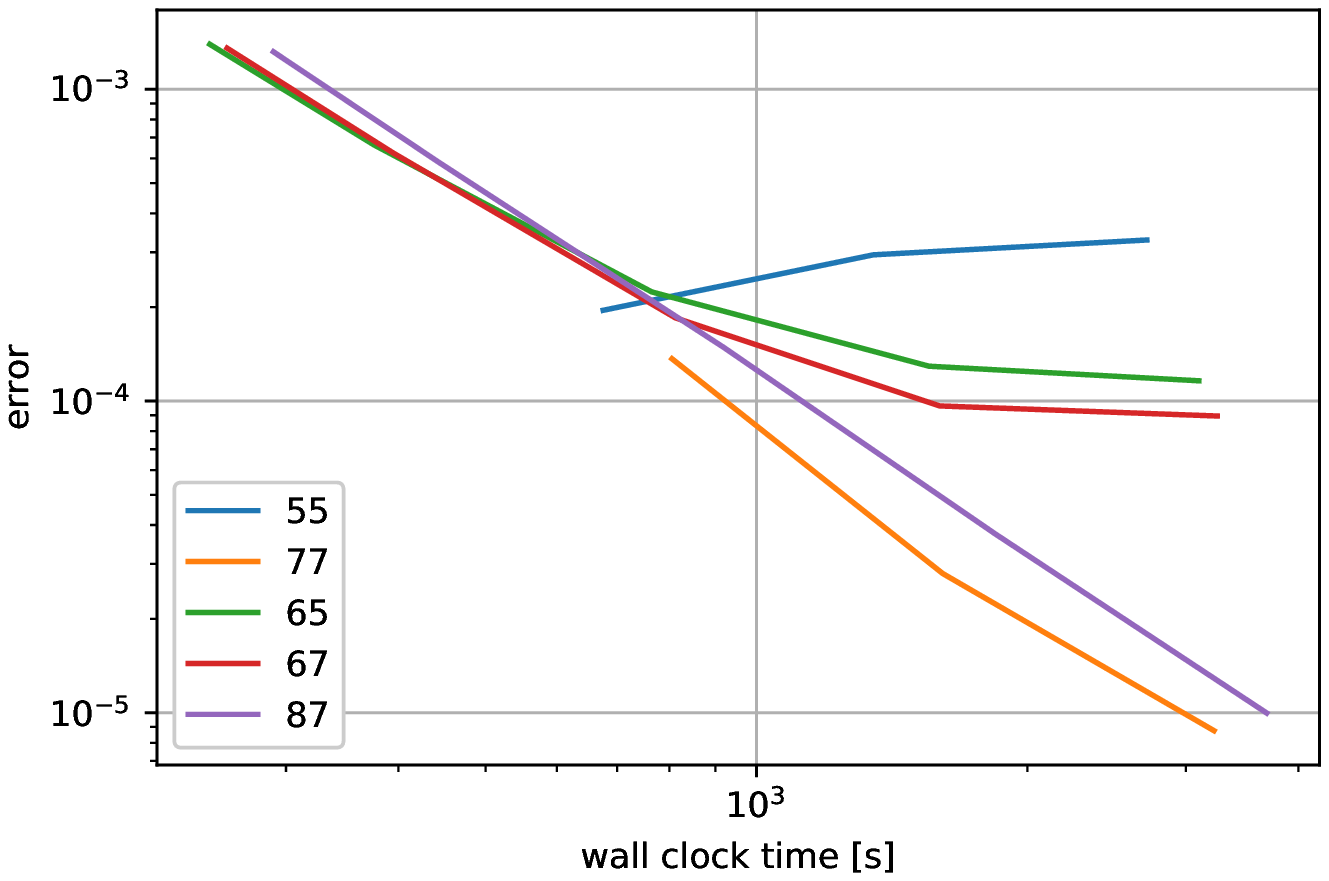}
\caption{Landau damping: Error in electrostatic energy as a function of the CPU time for various interpolators.  The numbers indicate the stencil widths of
the $\xb$ and the $\vb$ interpolations, respectively.}
\label{fig:landausum_cpu_error}
\end{figure}

Figure \ref{fig:landausum_cpu_error} shows the accuracy as a function of the total CPU time for a simulation on a single node of the DRACO cluster with 32 MPI processes for the Landau case. We can see that the computing time increases with the order but the increase is very small. On the other hand, the increasing halo cells for increasing order will impact the performance more strongly when MPI communication between nodes is involved.


\subsection{Simulations with rotating mesh}

As an example of a simulation with a strong background magnetic field $\Bb_0$, let us consider a simulation with initial value given as
\begin{equation}
f(\xb, \vb) = (1+\alpha \cos(k_{\perp}x_1) \cos(k_{\parallel}x_3) ) \exp\left(-\frac{\|v\|^2}{2}\right).
\end{equation}
In this case, the gyrofrequency is given by $\omega_c = \frac{2\pi}{B}$.

In order to understand the time-step restrictions of our distributed memory parallelization with the rotating grid, we estimate the maximum displacement in the various advection steps. The displacement of an $x_1$ advection at time $t$ is given by
\begin{equation}
\begin{aligned}
& |\Delta t \left( \cos(Bt) v_{1} + \sin(Bt) v_{2} \right)| \\
& \leq \Delta t \left( |\cos(Bt)| v_{1,\max} + |\sin(Bt)| v_{2,\max} \right) \\
& \leq \Delta t \sqrt{2} \max( v_{1,\max}, v_{2,\max}).
\end{aligned}
\end{equation}
Here, we use $v_{i,\max}$ to denote the (in modulus) largest value of the velocity on the computational grid. For the displacement of the $x_2$ advection, the same estimate can be derived. The displacement of the $x_3$ advection is given by $\Delta t v_3$ and can thus be estimated by $v_{3,\max}\Delta t$.

The displacement of the velocity advections depends on the electric field. The electric field induced by the initial condition is given as
\begin{equation}
	\Eb = -\frac{\alpha}{k_{\perp}^2+k_{\parallel}^2} \begin{pmatrix}
	k_{\perp}\sin(k_{\perp}x_1) \cos(k_{\parallel}x_3) \\
	0 \\
	k_{\parallel}\cos(k_{\perp}x_1) \sin(k_{\parallel}x_3)
	\end{pmatrix}.
\end{equation}
If the electric field is damped in time, we can estimate the electric field by $\frac{\alpha k_{\perp}}{k_{\perp}^2+k_{\parallel}^2}$ for the perpendicular and $\frac{\alpha k_{\parallel}}{k_{\perp}^2+k_{\parallel}^2}$ for the parallel direction.

\begin{figure}
\includegraphics[width=\columnwidth]{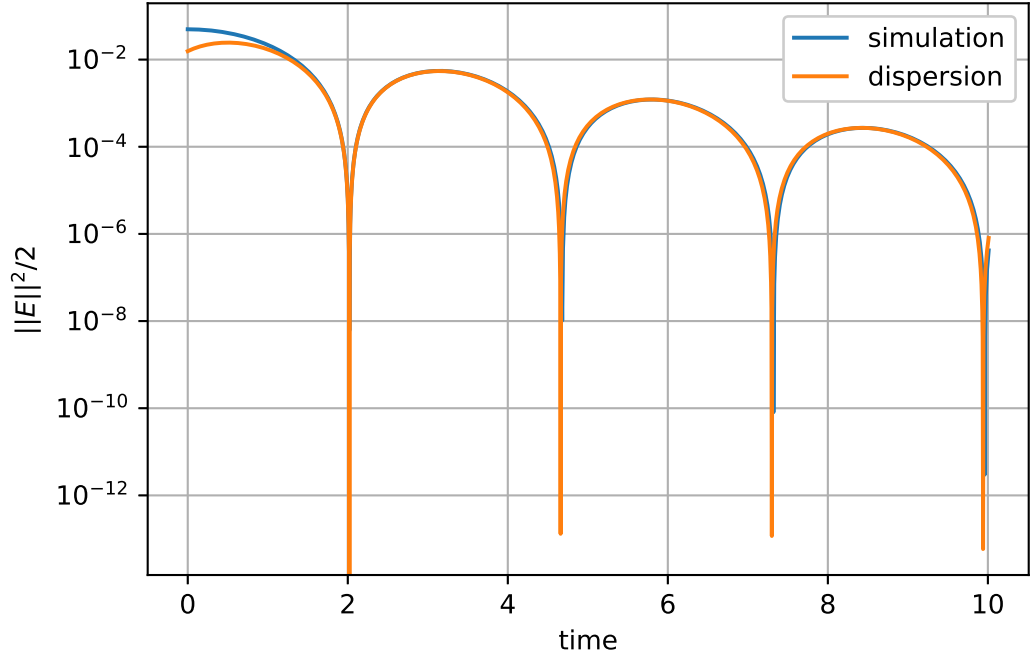}
\caption{Simulation with rotating velocity grid: Electrostatic energy simulated with $\Delta t = \omega_c/20$ and curve fitted from dispersion.}\label{fig:moving_dispersion}
\end{figure}

Let us consider the following parameters, $B=20\pi$ ($\omega_c=0.1$), $k_{\perp} = k_{\parallel} = 0.5$, $\alpha = 0.01$. Using 16 grid points along each spatial dimension and 32 points along the velocity dimensions as well as a velocity domain limited to $[-6,6]$, the grid spacing takes the values of $\Delta x_i = \frac{4\pi}{16} = \frac{\pi}{4}$ and $\Delta v_i = \frac{12}{32} = 0.375$. A linear dispersion relation gives $\omega = 1.1900 - 0.2843\im$ as a coefficient in the temporal Laplace transform, i.e.~the electric field is damped. Figure \ref{fig:moving_dispersion} shows the solution with time step $\Delta t = \omega_c/20$ compared to the solution with frequency and damping rate predicted by the dispersion. We see that the simulation is in good agreement with the dispersion, except for the first oscillation as usual.

\begin{figure}
\includegraphics[width=\columnwidth]{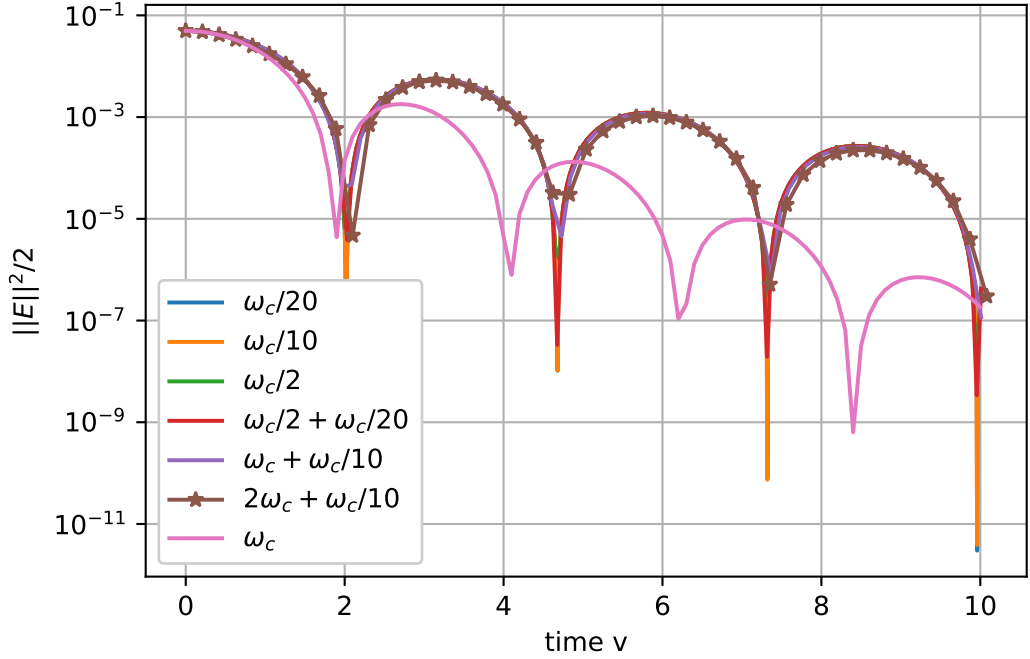}
\caption{Simulation with rotating velocity grid: Electrostatic energy for various time steps.}\label{fig:moving_dt}
\end{figure}

The maximum displacement along $x_1$ and $x_2$ direction is given by $\Delta t 6\sqrt{2}$ and hence the displacement is restricted to one cell size if $\Delta t \leq 9.256 \cdot 10^{-2}$, i.e. slightly below one period of the gyration. On the other hand, the displacement of the velocity advections is bounded by $0.01 \Delta t$ such that the displacement is smaller than the grid size for all $\Delta t \leq 37.5$. Figure \ref{fig:moving_dt} shows the electrostatic energy from simulations with various time steps. In the $\xb$ advection steps, we use centered Lagrange interpolation with 6 points and in the $\vb$ advections a fixed 7-point Lagrange interpolation formula. Note that we use symmetric halo cells of sufficient size for the $x_1, x_2$ directions since a static blocking as in the pure Vlasov--Poisson case is not possible on the rotating mesh. The results show that rather good results can be obtained for time steps above the gyrofrequency. In particular, we can use time steps close to $\omega_c/2$ which would yield a completely nonlocal stencil if we would not rotate the mesh. However, the time step cannot be a multiple of $\omega_c$ since then the magnetic field cancels out in the propagator on the rotating mesh.



%

%
%
%

\section{Performance benchmarks}\label{sec:performance}

\begin{table*}
\caption{Specification of the hardware used for performance evaluation.  On the
KNL node, only the fast on-chip MCDRAM was used as indicated by the underline.}
\label{tab:nodes}
\centering
{\tiny
\begin{tabular}{lllll}
\hline System & SuperMUC & DRACO & KNL node  & Marconi-KNL\\
\hline CPU & Intel SandyBridge & Intel Haswell & Intel Knights Landing & Intel Knights Landing\\
           & 2 $\times$ Xeon E5-2680 & 2 $\times$ Xeon E5-2698v3 & Xeon Phi 7210 & Xeon Phi 7250 \\
Cores     & $2 \times $8 & $2 \times $16 & 64 & 68\\
Threads   & up to 2 per core & up to 2 per core & up to 4 per core &up to 4 per core\\
Frequency & 2.7 GHz & 2.3 GHz & 1.3 GHz & 1.4 GHz\\
Memory    & 2 $\times$ 16 GB (50 GB/s) & 2 $\times$ 64 GB (68 GB/s) &
\underline{16 GB MCDRAM} (450 GB/s) &  16 GB MCDRAM\\
& & & 96 GB (90 GB/s) &  96 GB of DDR4\\
SIMD      & AVX & AVX2 & AVX-512 &  AVX-512\\
Network   & Mellanox FDR10 (40 GB/s) & M'x FDR14 (56 GB/s) & --- &  Intel OmniPath (100 GB/s) \\
\hline \end{tabular}
}
\end{table*}
To systematically compare and evaluate the performance and the scalability of
the remap and the decomposition implementations a series of
runs was performed, going from a single compute node to a HPC cluster, and
further up to multiple islands on a supercomputer.
For most of our basic tests and during iterative performance optimization work,
we used a two-socket Intel Haswell-type node, the building block of the DRACO
HPC cluster at 
MPCDF\footnote{DRACO HPC extension,
\url{http://www.mpcdf.mpg.de/services/computing/draco}, accessed on 2017-12-13.}.
Moreover an Intel Xeon Phi KNL node was available, running in ``flat'' mode with
the 16 GB of MCDRAM available as a separate memory domain.  All the runs
performed on the KNL used the fast MCDRAM exclusively.
Finally, we present large-scale runs performed on the SandyBridge partition of
the SuperMUC HPC system of the Leibnitz Supercomputing Centre, covering up to 8
islands with 64k physical cores.
Table \ref{tab:nodes} provides details on the specifications of the compute
nodes.

The Landau damping test case was chosen in each run. We consider a 7-point
Lagrange interpolation for both $\xb$ and $\vb$ advections since it proved to be
a good choice in terms of accuracy and flexibility according to the numerical comparison
presented in the previous section. Note that the 6-point centered Lagrange interpolation has the same halo width and therefore shows a similar performance.
Any wall clock time given refers to the computation of 5 time steps.
The initial time step is excluded from the
time measurement to compensate for the initialization overhead of the MPI
library which can be significant.

To compile and link the code, recent versions (16 and 17) of the Intel compiler
were used throughout this work, with the optimization flags set to
\texttt{-O3 -xHost -ipo-separate -qopenmp}.  On DRACO and on the Xeon Phi node,
Intel MPI is used whereas on SuperMUC, IBM PE is used.  We first look at the
performance on a single node, focussing on both process (MPI) and thread
(OpenMP) parallelism.

\subsection{Node-level performance}

To evaluate the performance of the OpenMP-based parallelization in comparison to
MPI we present scalings on a single compute node in this section.

Each run was repeated 5 times to average out small variations between individual
runs which were found to be on the order of up to 5\%.  A problem size
of $32^6$ was chosen since it is quadratic and represents a reasonable (though
moderately large) size for a per-socket workload on the relevant systems, the
net size of the distribution function being 8 GB in double precision.

Moreover, the simulation of a $32^6$ hypercube fits completely into the
high-bandwidth memory of the KNL node, at least for the domain decomposition
implementation considered mainly in this paper.
The same or a comparable size for the per-socket workload will be used for the
large-scale runs presented below.
The aim of this section is to show that the domain decomposition implementation
delivers excellent parallel performance on a single node in both MPI and OpenMP.
This feature paves the way to efficient hybrid-parallel large-scale simulations
performed on many-core distributed-memory HPC clusters.  Benchmark runs will be
discussed in the next section.

\begin{figure}[htbp] \centering
\includegraphics[width=\columnwidth]{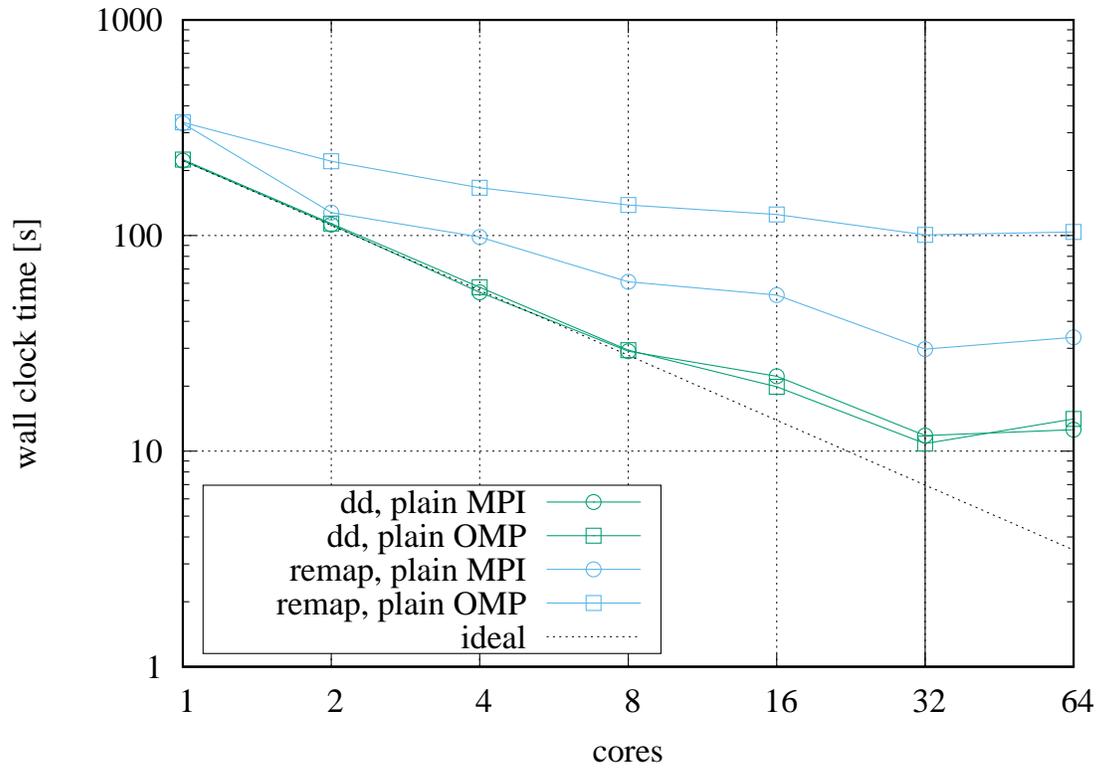} \caption{Strong scaling in
pure MPI and pure OpenMP on the Haswell node, comparing the domain decomposition code with the remap implementation, running 5 time steps of
a $32^6$ case with 7-point Lagrange interpolation.  The vertical
line indicates the transition to simultaneous multithreading.}
\label{fig:hsw_scan}
\end{figure}

Figure \ref{fig:hsw_scan} shows strong scalings on the Haswell-type node.
For both, the domain decomposition and the remap case, a scan in the number of
MPI tasks keeping the number of threads fixed to 1, and for comparison, a scan
in the number of OpenMP threads with a single MPI process are shown.

For each series the optimum pinning strategy is chosen, which has been
determined experimentally a priori.  The MPI processes are pinned in a
round-robin fashion between the two sockets.  Doing so the memory bandwidth of
both sockets is used as soon as more than one process is run.  Only MPI
messages are transferred via the link between the sockets.
On the other hand, the threads for the OpenMP scan are pinned in a compact
fashion to physical cores, filling the first socket completely before going to
the second socket.  The compact OpenMP thread pinning strategy reduces the
traffic over the link between the two NUMA domains significantly.  Note that a
thread-aware first-touch memory allocation and handling is not possible when
traversing the 6d array of the distribution function in all the 6 directions.

Fig.~\ref{fig:hsw_scan} shows that for the domain decomposition runs both the
MPI and OpenMP results are very similar and scale virtually ideally up 8 cores.
They continue to scale well up to the full $32$ physical cores of the node.
Adding hyperthreads does not improve either case, the wall clock times even
rise marginally, indicating that the CPU pipelines are already used efficiently.

The remap code performs significantly worse than the domain decomposition
implementation.  Running on the full node with 32 MPI processes it is about a
factor of $2.5$ slower.  Moreover it scales worse, achieving a speedup in MPI of
about $11.2$ compared to the speedup of about $18.8$ for the domain
decomposition code.  In OpenMP the remap code does not scale well which is due
to the simple cause that the remap operation takes place within the MPI library
as a (trivial) all-to-all operation which is inherently not threaded.
To have a fair comparison, the latter finding is one of the main reasons why we
mostly present pure MPI or only weakly-hybrid (using 2 hyperthreads per process,
for example) runs in the sections below on medium- and large-scale runs when we
compare the remap to the domain decomposition codes.


\begin{figure}[htbp] \centering
\includegraphics[width=\columnwidth]{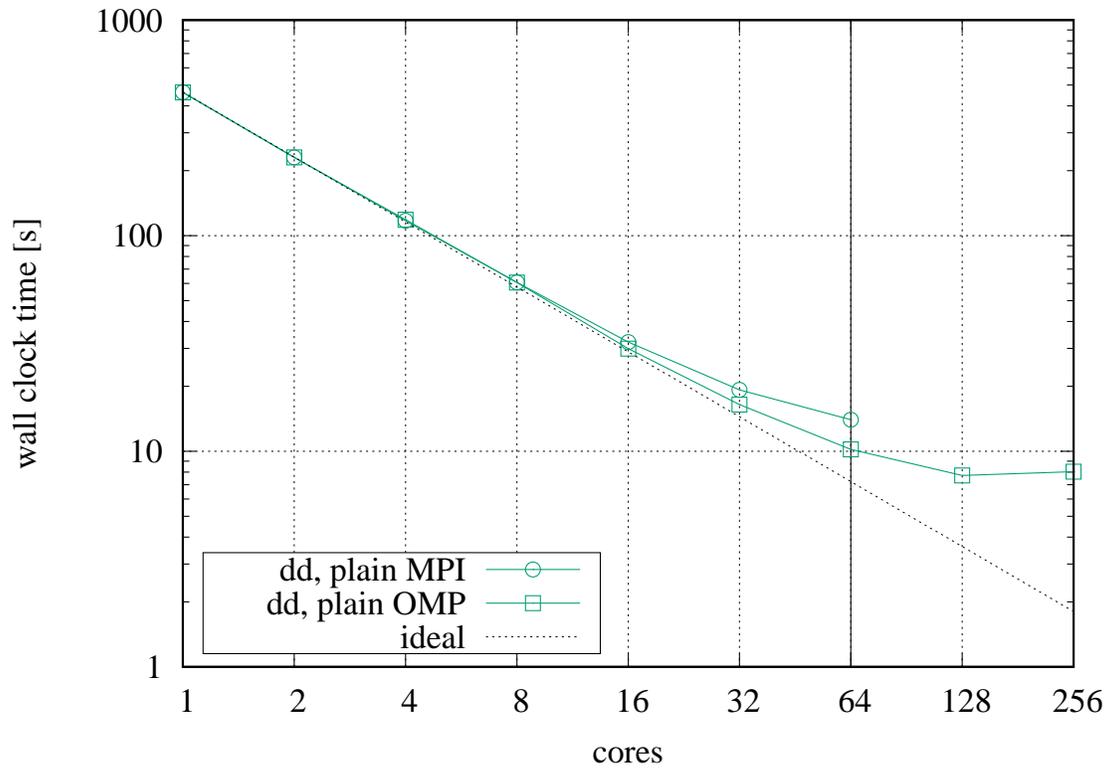}
\caption{Strong scaling in pure MPI and pure OpenMP on the Xeon Phi node.  For
comparison, the same setup as as the one used for Figure \ref{fig:hsw_scan} is
run.  The vertical line indicates the transition to simultaneous
multithreading.}
\label{fig:knl_scan} \end{figure}
For comparison and to include results from a many-core platform, we show numbers
based on the identical $32^6$ test case from domain decomposition runs performed
on the Xeon Phi node.  As seen in Fig.~\ref{fig:knl_scan},  the strong scaling
curves follow an ideal scaling up to 16 cores in both MPI and OpenMP, and
continue to scale well going up to the 64 available physical cores.  Scaling
further into the hyperthreading regime gives some benefit with 2 threads per
core, going further to 4 threads per core the performance even deteriorates
slightly.  The speedups reported here are approximately $33.0$ for the best MPI
case on 64 cores, and approximately $59.7$ for the best OpenMP case running on
128 hyperthreads.
Note at this point that the remap code requires more than 16 GB of memory in
the given setup and, hence, could not be considered on the KNL node.  Given the
result on the Haswell node it is not likely to outperform the domain
decomposition code on the KNL.  For similar reasons, we could not run more than
64 MPI processes using hyperthreads which does not make sense from a technical
point of view either.
%
%
While the plain MPI and plain OpenMP strong scalings are well-suited tools to
probe the parallelization efficiency individually, the HPC application user is
interested in the smallest time-to-solution which typically requires to choose a
mix of both the parallelization strategies at runtime which we focus on next.

\begin{table}\caption{Overview on the node-level performance by listing the 10
fastest runs on each platform. The use of hyperthreading is indicated by an
asterisk.} \label{tab:knl_vs_hsw} \centering \begin{tabular}{rrr|rrr} \hline
\multicolumn{3}{c|}{Haswell
node} & \multicolumn{3}{c}{Xeon Phi node} \\ \hline MPI & OpenMP & time [s] & MPI &
OpenMP & time [s] \\ \hline
1 & 32 & 10.84          & 1 & \mbox{*}128 & 7.73 \\
16 & \mbox{*}4 & 11.13  & 1 & \mbox{*}256 & 8.04 \\
16 & 2 & 11.40          & 1 & 64 & 10.21 \\
32 & \mbox{*}2 & 11.51  & 2 & \mbox{*}128 & 10.61 \\
32 & 1 & 11.80          & 8 & \mbox{*}32 & 11.11 \\
8 & \mbox{*}8 & 11.98   & 32 & \mbox{*}8 & 11.22 \\
8 & 4 & 12.05           & 16 & \mbox{*}16 & 11.48 \\
64 & \mbox{*}1 & 12.59  & 64 & \mbox{*}4 & 11.52 \\
1 & \mbox{*}64 & 12.81  & 4 & \mbox{*}64 & 11.91 \\
4 & 8 & 13.08           & 64 & \mbox{*}2 & 12.06 \\
\end{tabular}  \end{table}
Table \ref{tab:knl_vs_hsw} compares the node-level performance of the domain
decomposition runs between the Haswell and the Xeon Phi node.
%
%
Merely based on the ranked timings, the performance advantage of the KNL node
relative to the HSW node is roughly $1.4$.  Comparing the fastest runs with more
than one MPI process---which would be a practically more relevant case---the
advantage of the KNL melts down to only a factor of $1.05$.



In summary, both, the distributed-memory and the shared-memory parallelization
of the domain decomposition implementation were found to scale very well on a
single node.  OpenMP threads and MPI processes can virtually be used
interchangeably when running the domain decomposition implementation.  This is
an important finding relevant to the larger setups on many compute nodes that
rely on hybrid parallelization, e.g. running one process per socket and keeping
all the available cores busy with threads, as presented in the following
section.
However, in practice for a given setup, it may not be beneficial to arbitrarily
swap threads and processes due to implicit changes in the partitioning of the
numerical grid and the process grid which may taint the performance in some
cases.

\subsection{Parallel performance}

\subsubsection{Medium-scale runs.}

In this section, we focus on the parallel performance of the 6d Vlasov code on
distributed-memory HPC clusters.  We mainly consider the domain decomposition
implementation but draw some comparison to the remap code as well.

First we discuss medium-sized scaling runs performed on the DRACO HPC machine at
MPCDF, mainly to investigate the scalability of the hybrid implementation.  It
consists of Haswell nodes as detailed in Table \ref{tab:nodes} in the
previous section.  The nodes are grouped into islands of 32 machines and are
connected via an FDR14 InfiniBand network.  The maximum job size is 32 nodes
with 1024 physical cores in total, supporting up to 2048 simultaneously active
threads.
%
%
Note that the minimum number of MPI processes for which the domain decomposition
implementation communicates with all the process neighbors via MPI messages is
64 which is therefore taken as the baseline for all scalings.

To be able to include the domain decomposition code with pipelining, i.e.
overlapping of computation and communication, into the comparisons in a fair
way, we allow each process to use two hyperthreads even in a plain MPI run,
since the pipelining scheme relies on having multiple active threads per process
to work properly.  For the other two codes under consideration the effect of the
two hyperthreads is negligible as shown in the previous section on the
single-node performance.

\begin{figure}
\includegraphics[width=\columnwidth]{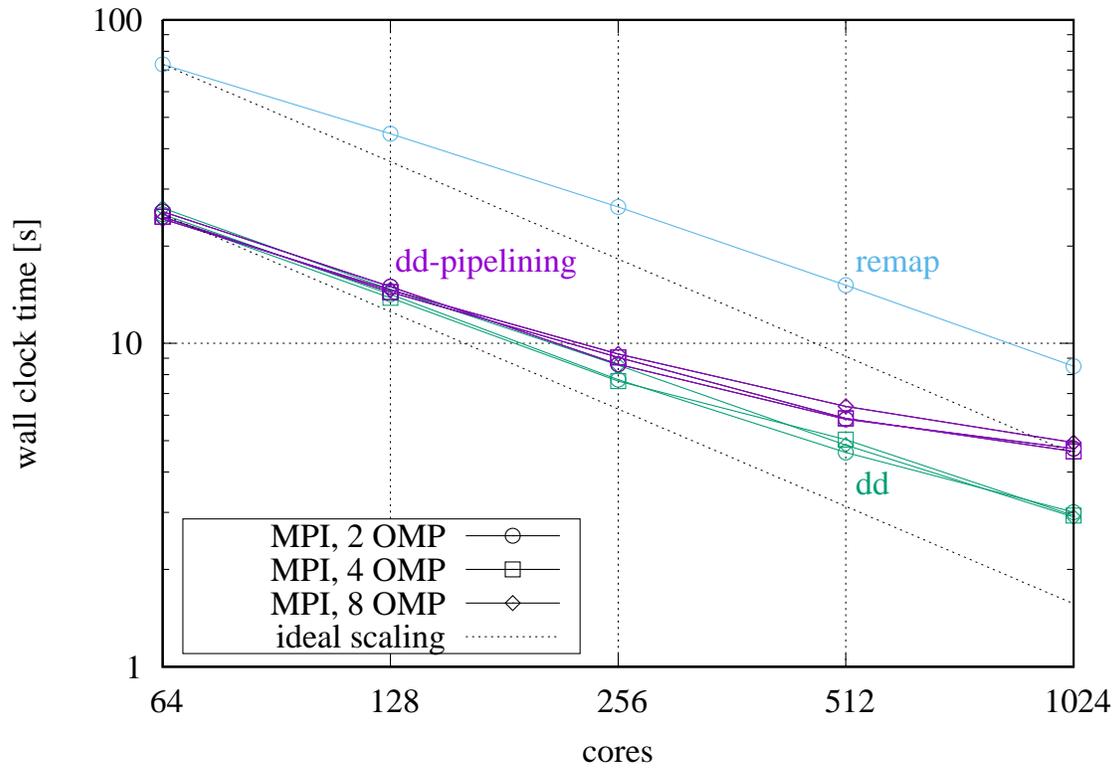}
\caption{Strong scaling on DRACO using 7pt Lagrange interpolation, running a
resolution of $32^4 64^2$, i.e. starting at a problem size of $32^6$ points per
CPU socket on two nodes.  The domain decomposition implementations with and
without pipelining are compared to the remap implementation. For the
measurement, 5 time steps are run, after discarding a first one to avoid MPI
initialization overhead.}
\label{fig:draco_mpi_strong}
\end{figure}
We now turn towards a strong scaling before the more relevant weak scaling
scenario is considered in greater detail.
Fig.~\ref{fig:draco_mpi_strong} shows strong scalings in plain MPI and also for
hybrid setups, going from 64 cores on two nodes up to the full island
size of 32 nodes.  The test setup uses a resolution of $32^4 64^2$, thereby
following up on the resolution of $32^6$ per CPU socket as previously used for
the single-node tests.
The remap implementation is significantly slower than both the domain
decomposition codes, at the same time it scales best by achieving a speedup of
about 8.6 of 16, closely followed by the domain decomposition code at about 8.3.
On the full island, i.e.~on 32 nodes, the domain decomposition code is faster by
a factor of about $2.8$ compared to the remap code, a result very similar to the
outcome on the full node.
Initially starting out at virtually the same wall clock time as the domain
decomposition case, the pipelining code scales worse, illustrating the overhead of
the pipelining scheme when decreasing the workload per MPI process.

The hybrid parallelization turns out to be efficient when going beyond a single
compute node as well.
Fig.~\ref{fig:draco_mpi_strong} shows strong scalings for hybrid runs of the
domain decomposition codes in direct comparison to the MPI runs, reducing the
number of MPI processes by a factor of 1/2 or 1/4 and increasing the number of
threads by the inverse factor.  Note that hyperthreads are used in the example,
i.e. two threads share the same physical core.
As can be seen, the hybrid curves follow the plain MPI curves very closely.  For
the largest runs the hybrid case with 8 threads per MPI process turns out
to perform best for the domain decomposition code without pipelining.

An important effect induced by the multidimensional domain decomposition has to
be recalled at this point to explain the deterioration of the strong scaling.
As the global problem size is kept constant, the fraction of the distribution
function that has to be communicated between neighbors increases
as the strong scaling is performed due to the decrease in local grid size.  This
effect can be mitigated to some degree by performing hybrid runs.
On the other hand, for the remap implementation the total volume of communicated
data stays constant, whereas the number of MPI messages increases quadratically
with the number of processes.

These strong scaling curves are mainly given for reasons of completeness. As
high memory demands in combination with low arithmetic intensity are the main
challenges to be tackled when dealing with the 6d Vlasov-Poisson problem we turn
towards the weak scaling properties of the codes which are more relevant to
Physics simulations on large grids.

\begin{figure}
\includegraphics[width=\columnwidth]{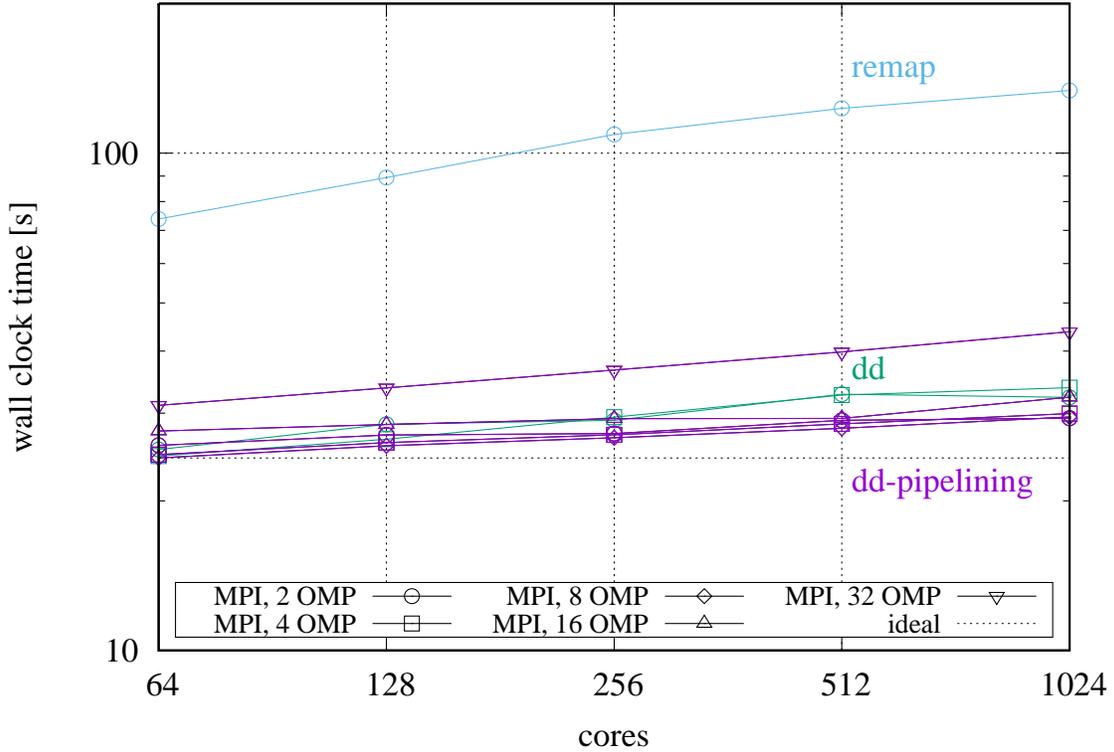}
\caption{Weak scaling on DRACO using 7pt Lagrange interpolation, keeping a local
problem size of $32^6$ points per CPU socket. Note that the largest runs have a
global problem size of $64^6$ points for the distribution function.}
\label{fig:draco_hybrid_weak_times}
\end{figure}
Fig.~\ref{fig:draco_hybrid_weak_times} shows a weak scaling, starting on two
nodes and keeping a workload of $32^6$ points per CPU socket.
Looking at the plain MPI runs (with 2 hyperthreads per process to enable
comparison to the pipelining implementation), the remap code is significantly
outperformed by the domain decomposition codes, with the pipelining
implementation being the fastest and scaling best.
The parallel efficiencies are approximately $0.88$ for the pipelining
implementation, $0.79$ for the non-pipelining one, and only $0.55$ for the remap
code.  On the full island, the pipelining code is about a factor of $4.6$ faster
than the remap code.

In addition to the plain MPI cases, the plot shows a selection of hybrid cases,
exchanging processes with threads, but operating on the same workload for
comparability.
The graph yields the important finding that hybrid setups running 4 or 8 threads
per process may be in fact the best choice for the pipelining code, which is
in this situation able to utilize several threads for each the advection and the
buffering tasks in parallel, while performing MPI communication at the same time,
thereby hiding parts of the communication behind useful computation.
As can be seen in addition for the pipelining implementation, the hybrid
parallelization performs very well over a broad range of configurations.  Even
running only two MPI processes per socket (labeled ``16 tpp'') turns out to be
very close to the case with one MPI process per physical core (labeled ``2
tpp''), less aggressive hybrid setups being located in between.  However,
reducing further to only a single process per socket (labeled ``32 tpp'') causes
the time to solution increase significantly which is caused by the fact that a
single simultaneous MPI operation per socket is not sufficient to saturate the
network interface.
Finally, it has to be recalled that each hybrid configuration uses a different,
optimal, process grid on top of the same numerical problem setup, which leads to
different communication patterns that contribute as well to the variation
observed between plain MPI and hybrid runs.

%
Concerning the weak scaling, the rise in the wall clock time is attributed to a
significant part to the fact that halos in more and more spatial directions are
communicated via the network instead of being shared via the memory on the node,
as the system size is increased.
An illustration and in-depth explanation will be given in the following section
dealing with scalings on SuperMUC using a component-wise breakdown of the run
time.

\begin{figure}[htbp]
\includegraphics[width=\columnwidth]{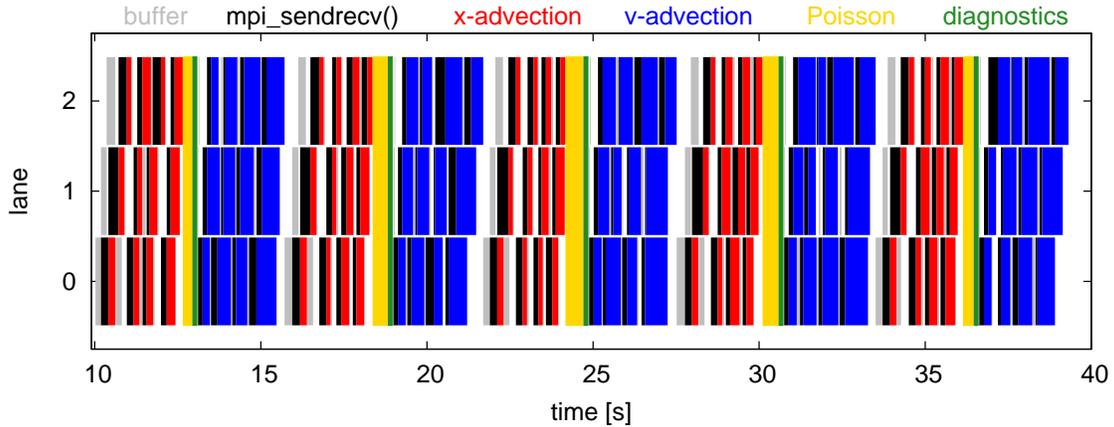}
\caption{Timeline view on a hybrid pipelining case, running 64 MPI processes on 16
nodes, with two MPI processes and 16 hyperthreads per socket, picked from the
runs shown in Fig.~\ref{fig:draco_hybrid_weak_times}.}
\label{fig:tasks_double}
\end{figure}
To illustrate the behavior and performance of the pipeline implementation, time
traces in a task-style are presented in Fig.~\ref{fig:tasks_double}.  In
Sec.~\ref{sec:pipelining},  the pipelining scheme was explained by means of the
schematic Fig.~\ref{fig:task_graph} which is hereby complemented with timing
data.
The run under consideration uses 64 MPI processes in total, employs two
processes per node, one per socket, and uses OpenMP threads to keep all the
cores busy.  The plot shows timings obtained from the control loop which
implements the pipelining scheme using three lightweight threads, indicated here
by three lanes. Temporal overlap is possible between the lanes within the $\xb$
advection and the  $\vb$ advection blocks where the number of nested threads for
each task is set such that at most the number of available hardware threads is
utilized.
As can be seen from Fig.~\ref{fig:tasks_double} the communication operations
shown in black overlap quite well with useful computation being performed at the
same time, though some minor jitter is present.
The fact that the $\vb$ advections and the $\xb$ advections cannot be overlapped due to
their blocking along different dimensions becomes visible from empty lanes when
the transition from $\vb$ (blue) to $\xb$ (red) occurs.
The configuration of the process grid in this example is $2^6$, with a local
numerical grid of $16 \cdot 32^5$ and a global grid of $32 \cdot 64^5$.  This
implies that there is more communication going on along the first dimension
which can be seen well from the first four black blocks of each $\xb$ advection
phase in Fig.~\ref{fig:tasks_double} which turn out to be wider than the rest.
Overall, the computations of the $\vb$ advections (blue) are more expensive than
the computation of the $\xb$ advections which is due to strided memory accesses,
as was already shown on Fig.~\ref{fig:single_core_clocks}. Note also that the
time spent in the Poisson step is comparably large and can vary quite a bit. The
reason is that it contains the reduction step which requires the synchronization
of comparably many processes and, hence, potential imbalances due to system
jitter during the advection steps are included in the Poisson timing.


In the following section, we discuss runs on a supercomputer at large scale,
also shedding more light on the effects which limit the parallel scalability.

\subsubsection{Large-scale runs.}

To further evaluate the performance of the domain-decomposition-based solver, we
present runs performed on the SuperMUC HPC system of the Leibnitz Supercomputing
Center\footnote{
SuperMUC Petascale System,
\url{https://www.lrz.de/services/compute/supermuc/systemdescription/}, accessed
on 2017-05-10.}.

On the so-called phase 1 partition of the SuperMUC system, each node is equipped
with two Intel E5-2680 CPUs (SandyBridge-EP) with 8 physical cores, each
supporting two threads per core (cf.~Table \ref{tab:nodes}).  There are 512
nodes per island, connected via an InfiniBand FDR10 network.  The blocking
factor of the network between islands is 1:4. In total, 18 islands are
available.  In the following, we present scalings going up to 8 islands with 64k
physical cores and 128k hardware threads.

The resolution of the test case considered for the weak scaling on SuperMUC is
$32^4 \cdot 16^2$ per socket, which is chosen---for reasons of the available
memory per node---a factor of four smaller than the size used per Haswell socket
on DRACO.  Note that the setup implies a local resolution of a $16^6$ hypercube
per process for plain MPI runs with 8 processes per 8-core socket, as shown
below in Table \ref{fig:mucbreakdown} which details on the grid parameters
involved in the scaling.  Starting from 64 processes on 4 nodes guarantees that
MPI communication takes place in all the 6 dimensions initially.

\begin{figure}
\includegraphics[width=\columnwidth]{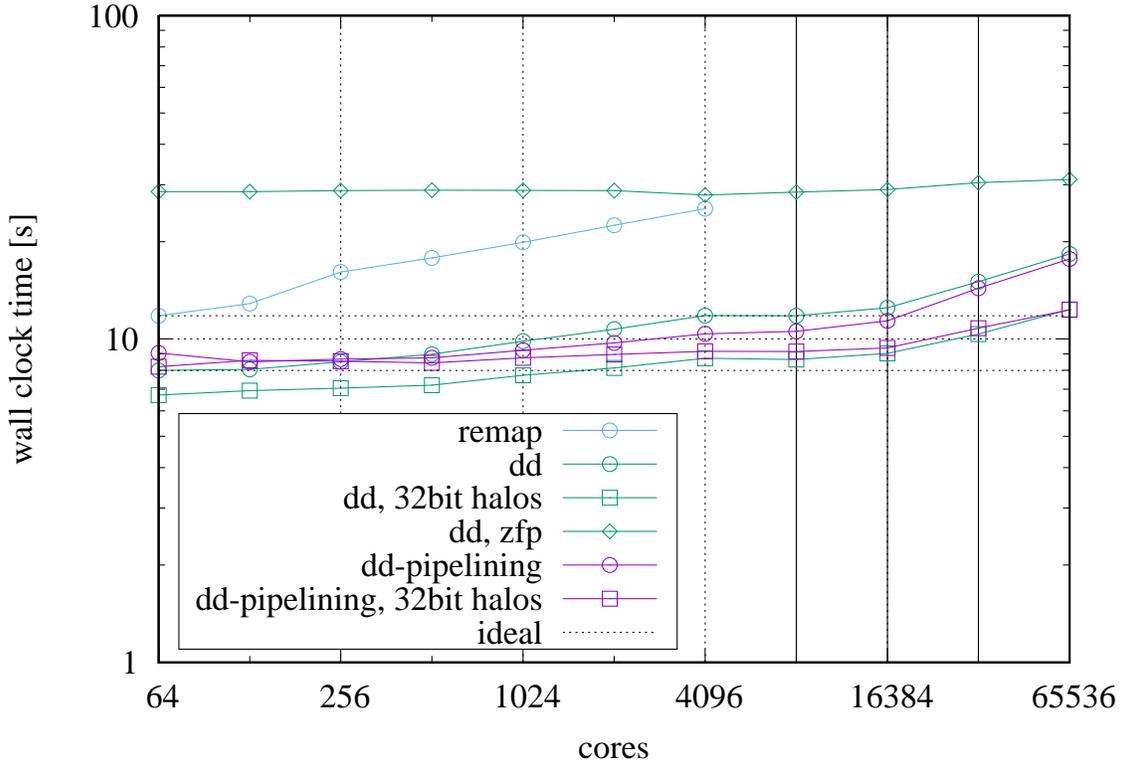}
\caption{Weak scaling on SuperMUC comparing the domain decomposition codes with
and without pipelining to the remap implementation, running one MPI process per
physical core, with 2 hyperthreads each.  The transitions going from 1 to 2, 2 to
4, and 4 to 8 islands are indicated by vertical lines.}
\label{fig:supermuc_mpi_weak}
\end{figure}
Fig.~\ref{fig:supermuc_mpi_weak} shows weak scalings for the remap and the two
domain decomposition codes.  There is one MPI process per physical core with 2
hyperthreads each.  These extra hardware threads are enabled in order to include
the pipelining implementation into the comparison.

The results show that the domain decomposition codes both scale quite well.  The plain
domain decomposition implementation turns out to be faster than the pipelining
one up to 512 MPI processes when the latter takes over.  Going beyond the island
boundaries leads to an increase in the wall clock times which is moderate at two
islands but becomes more significant when going to 4 or 8 islands.
When scaling up, the increase in the wall clock time is mainly caused by the
fact that an increasingly larger fraction of the halo data is exchanged over the
network between nodes and finally between islands as the problem size is
increased.  A run time analysis of the building blocks of the code is discussed
below and sheds more light on this issue.

Compared to the domain decomposition codes, the remap code is slower, as already
known from the previous tests.  In addition it scales slightly worse.  More
details on its scalability limitations will be given below.  Due to a restriction
in the implementation, the Poisson step currently prevents to use more than 4k
MPI processes.

In addition, the weak scaling plot Fig.~\ref{fig:supermuc_mpi_weak} contains
data from two experiments aimed at increasing the parallel scalability by
reducing the size of the MPI halo messages.  The first experiment (labeled
``32bit halos'') simply sends the halo data in single precision, thereby halving
the communication volume.  The wall clock time is significantly reduced and also
the scalability is improved to some degree, especially when going from one to
two islands.
The second experiment (labeled ``zfp'') takes one step further by applying a
lossy compression to the halo data, thereby effectively reducing it to about a
quarter of its size while preserving single precision accuracy.  This comes at a
significant computational cost but at the same time leads to a virtually flat
scaling with nearly 100\% parallel efficiency, even when crossing island barriers.
Interestingly, at 4096 MPI processes, the run time of the remap code is already
very close to the one of the domain decomposition code with ZFP compression
enabled.
As hardware standards evolve it is to be expected that the penalty of the ZFP
compression becomes less and less important for hybrid runs on present and
future systems with greater relative compute power compared to the network
bandwidth.  

Hybrid setups, i.e.~swapping MPI processes in favor of OpenMP threads for the
domain decomposition code, did not turn out to be better than plain MPI cases on
SuperMUC with its comparably small number of cores, in contrast to DRACO as
shown before.  As argued already for ZFP, the hybrid feature will become more
important in the future.

Let us now turn towards an in-depth analysis of the scaling of the building
blocks of the domain decomposition code when performing the weak scaling.
\begin{figure}
\includegraphics[width=\columnwidth]{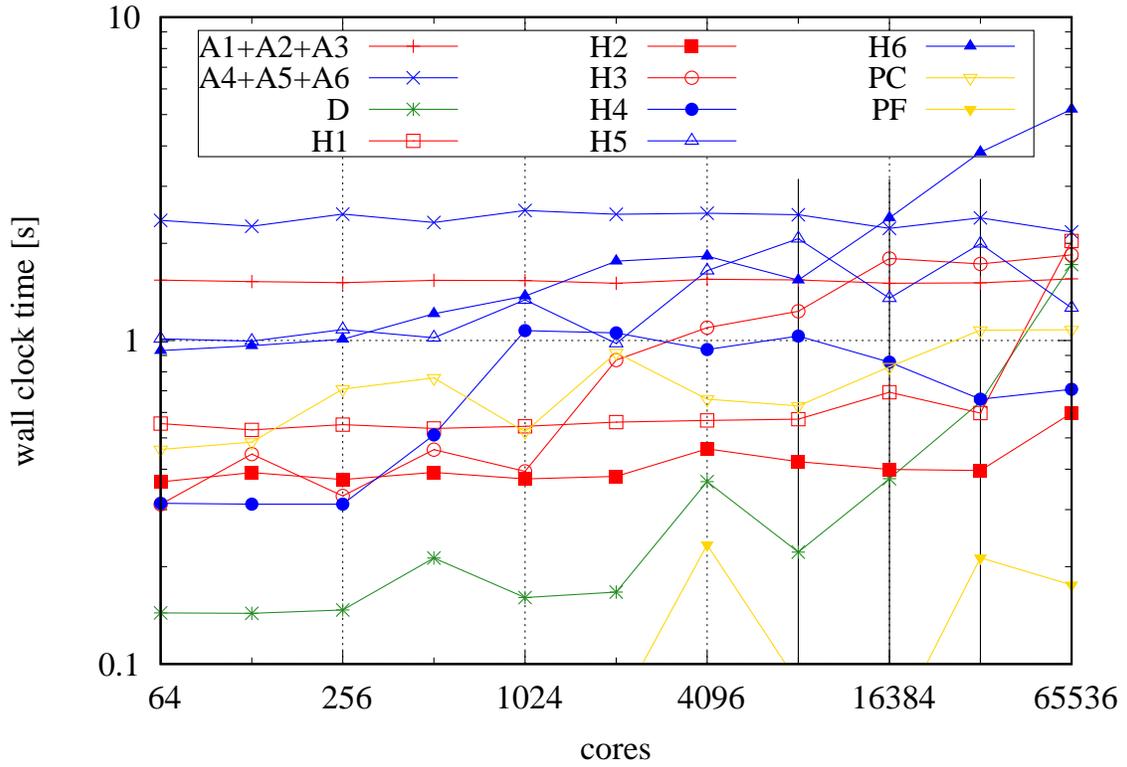}
\caption{Weak scaling of the building blocks of the implementation for the plain
MPI domain decomposition run shown in Fig.~\ref{fig:supermuc_mpi_weak}, scanning
from 64 up to 65536 MPI processes.
The letter A indicates the advection computations whereas H indicates the
halo-exchange operations, for both of which the dimension is given.
The letter D represents the diagnostics and P the Poisson solver.  In
particular, PC means the charge density computation whereas PF means the field
solver component, respectively.
The X advections and the V advections were grouped together.
The transitions going from 1 to 2, from 2 to 4, and from 4 to 8 islands are
indicated by vertical lines.}
\label{fig:mucbreakdown}
\end{figure}
\begin{table}[htbp]
\caption{Overview on the configurations of the various grids (global grid, MPI
process grid, local grid, distances between neighboring processes in units of
processes and in units of nodes) involved in the domain decomposition setups
of the weak scalings shown in Figs.~\ref{fig:supermuc_mpi_weak} and
\ref{fig:mucbreakdown}.}
\label{tab:breakdown_grid}
\centering
{\scriptsize
\begin{tabular}[t]{r|rrrrrr}
\hline
cores & x1 & x2 & x3 & x4 & x5 & x6 \\ \hline\hline
\multicolumn{7}{c}{global grid} \\ \hline
 64    & 32 & 32 & 32 & 32 & 32 & 32 \\
 128   & 32 & 32 & 32 & 32 & 32 & 64 \\
 256   & 32 & 32 & 32 & 32 & 64 & 64 \\
 512   & 32 & 32 & 32 & 64 & 64 & 64 \\
 1024  & 32 & 32 & 64 & 64 & 64 & 64 \\
 2048  & 32 & 64 & 64 & 64 & 64 & 64 \\
 4096  & 64 & 64 & 64 & 64 & 64 & 64 \\
 8192  & 64 & 64 & 64 & 64 & 64 & 128 \\
 16384 & 64 & 64 & 64 & 64 & 128 & 128 \\
 32768 & 64 & 64 & 64 & 128 & 128 & 128 \\
 65536 & 64 & 64 & 128 & 128 & 128 & 128 \\
\hline
\multicolumn{7}{c}{process grid} \\ \hline
 64    & 2 & 2 & 2 & 2 & 2 & 2 \\
 128   & 2 & 2 & 2 & 2 & 2 & 4 \\
 256   & 2 & 2 & 2 & 2 & 4 & 4 \\
 512   & 2 & 2 & 2 & 4 & 4 & 4 \\
 1024  & 2 & 2 & 4 & 4 & 4 & 4 \\
 2048  & 2 & 4 & 4 & 4 & 4 & 4 \\
 4096  & 4 & 4 & 4 & 4 & 4 & 4 \\
 8192  & 4 & 4 & 4 & 4 & 4 & 8 \\
 16384 & 4 & 4 & 4 & 4 & 8 & 8 \\
 32768 & 4 & 4 & 4 & 8 & 8 & 8 \\
 65536 & 4 & 4 & 8 & 8 & 8 & 8 \\
\hline
\multicolumn{7}{c}{local grid} \\ \hline
any    & 16 & 16 & 16 & 16 & 16 & 16 \\
\hline
\multicolumn{7}{c}{distance in processes} \\ \hline
 64    & 1 & 2 & 4 & 8 & 16 & 32 \\
 128   & 1 & 2 & 4 & 8 & 16 & 32 \\
 256   & 1 & 2 & 4 & 8 & 16 & 64 \\
 512   & 1 & 2 & 4 & 8 & 32 & 128 \\
 1024  & 1 & 2 & 4 & 16 & 64 & 256 \\
 2048  & 1 & 2 & 8 & 32 & 128 & 512 \\
 4096  & 1 & 4 & 16 & 64 & 256 & 1024 \\
 8192  & 1 & 4 & 16 & 64 & 256 & 1024 \\
 16384 & 1 & 4 & 16 & 64 & 256 & 2048 \\
 32768 & 1 & 4 & 16 & 64 & 512 & 4096 \\
 65536 & 1 & 4 & 16 & 128 & 1024 & 8192 \\
\hline
\multicolumn{7}{c}{distance in nodes} \\ \hline
64    & 0.1 & 0.1 & 0.3 & 0.5 & 1.0 & 2.0 \\
128   & 0.1 & 0.1 & 0.3 & 0.5 & 1.0 & 2.0 \\
256   & 0.1 & 0.1 & 0.3 & 0.5 & 1.0 & 4.0 \\
512   & 0.1 & 0.1 & 0.3 & 0.5 & 2.0 & 8.0 \\
1024  & 0.1 & 0.1 & 0.3 & 1.0 & 4.0 & 16.0 \\
2048  & 0.1 & 0.1 & 0.5 & 2.0 & 8.0 & 32.0 \\
4096  & 0.1 & 0.3 & 1.0 & 4.0 & 16.0 & 64.0 \\
8192  & 0.1 & 0.3 & 1.0 & 4.0 & 16.0 & 64.0 \\
16384 & 0.1 & 0.3 & 1.0 & 4.0 & 16.0 & 128.0 \\
32768 & 0.1 & 0.3 & 1.0 & 4.0 & 32.0 & 256.0 \\
65536 & 0.1 & 0.3 & 1.0 & 8.0 & 64.0 & 512.0 \\
\hline
\end{tabular}
}
\end{table}
Fig.~\ref{fig:mucbreakdown} shows a breakdown of the individual components of
the implementation without overlap of communication and computation.  Table
\ref{tab:breakdown_grid} complements Fig.~\ref{fig:mucbreakdown} with
information on the grid configuration of the runs, essential for a better
understanding of the behavior.

The advection computations (A1--A6) scale virtually ideally. However,
from the communication-intensive operations such as the halo exchanges and the
Poisson step we clearly see the effects of the network topology.  In the runs
performed, the grid of MPI processes is laid out in column-major order, i.e. MPI
processes are closest in the 1st dimension and are farthest apart in the 6th
dimension as shown in Table \ref{tab:breakdown_grid} for each problem
size of the weak scaling.
In particular, the table gives information on how far processes, that are
neighbors in the process grid in a logical sense, are separated on
the machine in units of processes and nodes, respectively.  Note that on the
machine and in combination with the configuration we are considering, there are
16 cores (and MPI processes) per node such that any $x_1$-$x_2$ plane in any run
discussed here fits into a node.  This means that halo exchanges in $x_1$ or
$x_2$ direction take place in shared memory via the MPI library.
In Fig.~\ref{fig:mucbreakdown} this fact is evident from the virtually ideal
scaling of the H1 and H2 curves.  The operation H2 is likely to be faster
because the batch system distributes the processes in a round-robin fashion between
the two sockets such that the H1 communication is done via the link between the two
NUMA domains whereas the H2 communication takes place within the domain.
On the contrary, neighbors in the 5th and 6th dimension communicate exclusively
over the network for all the runs under consideration, causing H5 and H6 to be a
rather costly operation.  As can be seen the  communication in $x_6$ is the main
reason for the slowdown when crossing island barriers.
The fraction of halo data communicated between islands rises from 25\% on 2
islands to 50\% on 4 islands and finally to 100\% on 8 islands, as the
``distance in nodes'' values in Table \ref{tab:breakdown_grid} indicate for the
$x_6$ communication.
The H3 and H4 curves show the same behavior for the transition from intra-node
to inter-node communication, as the rise at 256 and 1024 cores, respectively,
indicates.

The Poisson step is broken down in two parts, the reduction step (PC) and the actual solution of the Poisson problem (PF). The latter part PF is negligible and scales well (except for some fluctuations especially visible in this step due to the small numbers). The reduction step, on the other hand, shows a similar increase as the $\vb$ halo exchange steps since it includes communication along all the velocity dimensions.

The diagnostics computation D is initially negligible, however, going to
larger process counts its significance rises, especially when crossing the
island boundaries.  Note that the diagnostics is computed purely local to a
process, and only in a final step an array of 12 double precision numbers is
summed (reduced) globally to rank 0 via MPI, causing the increase in time
observed in Fig.~\ref{fig:mucbreakdown}.

Finally, it has to be pointed out that the runs could be performed only once,
and hence, some fluctuations are to be expected, e.g. in the D curve at 1k and
8k in Fig.~\ref{fig:mucbreakdown} (and not being present in the D curve in
Fig.~\ref{fig:mucbreakdown_remap} below, for comparison).

\begin{figure}
\includegraphics[width=\columnwidth]{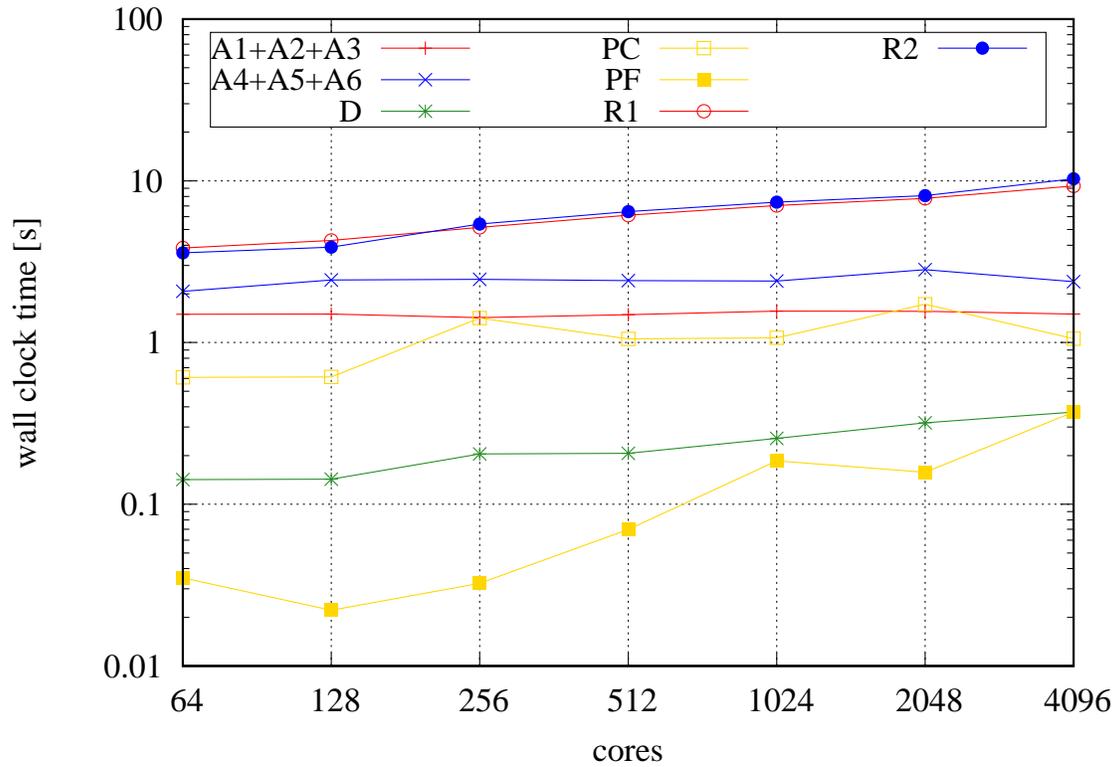}
\caption{Weak scaling of the building blocks of the implementation for the remap
run shown in Fig.~\ref{fig:supermuc_mpi_weak}.  The letter A indicates the
advection computations whereas R1 and R2 indicate the remap operations involving
all-to-all communication, with R1 mapping from the representation contiguous in
V to the representation contiguous in X, and R2 inversely.
The letter D represents the diagnostics and P the Poisson solver.  To
improve the clarity of the plot the X advections and the V advections were
grouped together.}
\label{fig:mucbreakdown_remap}
\end{figure}
For comparison, Fig.~\ref{fig:mucbreakdown_remap} shows the temporal breakdown
of the building blocks of the remap code.
Starting with 4 nodes (64 MPI processes) at a resolution of $32^6$ the remap
operations R1 and R2 dominate from the beginning over the compute-intense
advection computations, with their run time fraction steadily increasing in the
following.
The advection computations A1--A6 scale ideally as to be expected.  The Poisson
solve step turns out to take about the same time as the $\xb$ advection
computation, whereas the diagnostics is negligible in the range under
consideration.

\subsubsection{Intel manycore.}

\begin{figure}
\includegraphics[width=\columnwidth]{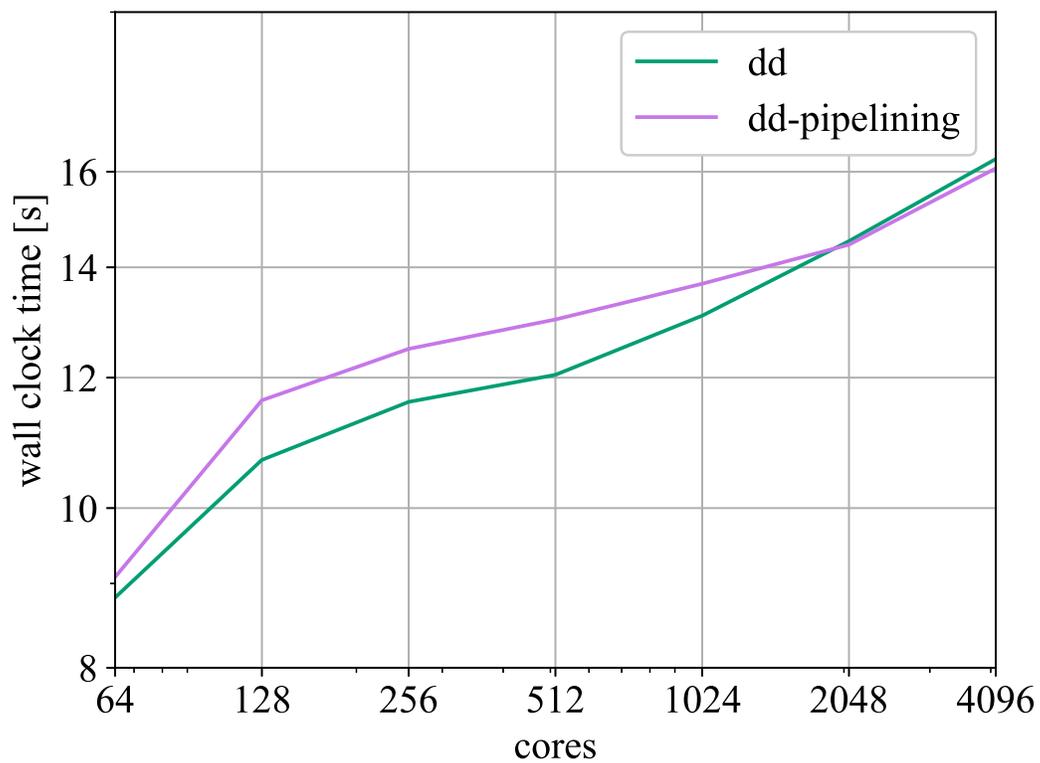}
\caption{Weak scaling on the Marconi KNL partition comparing the domain
decomposition codes with and without pipelining, running one MPI process per
physical core with 4 hyperthreads each.}
\label{fig:weak_scaling_knl}
\end{figure}

Finally, we have also run the same experiment on the Intel KNL partition of the
Marconi Fusion cluster at CINECA.  We report results from weak scaling starting
with a resolution of $32^6$ points on a single KNL node, distributed over 64 MPI
processes and allowing for 4 hyperthreads per MPI process. The configuration was
chosen such that the local grid is a regular hypercube of $16^6$ points per MPI
process. Fig.~\ref{fig:weak_scaling_knl} shows the wall clock time for the weak
scaling experiment, running the domain decomposition code with and without
pipelining.
A breakdown of the timings for the various steps shows a similar pattern as on
SuperMUC with virtually flat curves for the compute-only steps and increasing
communication times when intra-node communication is replaced by inter-node
communication. The H6 communication becomes expensive starting at 32 nodes (2048
cores).  At the same number of nodes, the pipelining scheme becomes beneficial
by hiding communication partly behind computation.
%

\section{Summary and Conclusions}\label{sec:conclusions}

In this paper, we have addressed the design, performance optimization, and
parallel scalability of a semi-Lagrangian Vlasov--Poisson solver in
six-dimensional phase space. The algorithm relies on dimensional splitting,
i.e.~consecutive interpolation along each of the dimensions.
We have addressed two parallelization schemes, the remapping method
and a 6d domain decomposition approach, the latter replacing the
all-to-all-type of communication of the former with a peer-to-peer (next
neighbor) communication pattern.

The domain decomposition method turns out to be superior to the remap method
with respect to memory, per-core and parallel performance. Parallel scalability
is demonstrated in a number of numerical experiments for various setups, going
up to 64k physical cores on a supercomputer.
By design, the domain decomposition uses halo cells which in practice introduces
restrictions on the time step. However, these are mitigated by using a one-sided
blocked communication scheme for the Vlasov--Poisson case and a rotating mesh
that follows a background magnetic field.
Using pipelined communication and computation the domain decomposition method
allows to efficiently overlap the communication with useful computation.

So far, we have only addressed Lagrange interpolation which suits the
distributed computations very well due to its locality. In the future, local
splines or discontinuous Galerkin interpolation schemes will also be considered.
Furthermore, extensions to Vlasov--Maxwell and more complex geometries are
natural enhancements. 
%
Due to its high efficiency and parallel scalability, the new code enables simulations in  six-dimensional phase space with grid sizes and resolutions large enough for
computing relevant physics cases and paves the way to systematically comparing gyrokinetic 
to fully kinetic simulations. 

\bibliographystyle{plainnat}

\bibliography{bsl6d}

\end{document}